\documentclass[aps,prb,reprint,superscriptaddress]{revtex4-2}
\usepackage{amsmath}
\usepackage{graphicx}
\usepackage{amsfonts}
\usepackage{color}
\usepackage{bm}
\usepackage{lipsum}
\usepackage{comment}

\usepackage{dsfont}
\usepackage{bm}
\usepackage{amsmath}
\usepackage{amssymb}
\usepackage{graphicx} 
\usepackage{braket}
\usepackage{mathtools}
\usepackage{times}
\usepackage{enumerate}

\newcommand{\ii}{\mathrm{i}}

\newcommand{\bea}{\begin{eqnarray}}
\newcommand{\eea}{\end{eqnarray}}

\newcommand{\Tr}{\mathrm{Tr}}

\newsavebox{\mstrut}
\newcommand{\bbra}[1]{%
    \sbox{\mstrut}{\(#1\)}%
    \mathinner{\langle\kern-0.3\ht\mstrut\left\langle{#1}\right|\mkern-2mu}%
}
\newcommand{\kett}[1]{%
    \sbox{\mstrut}{\(#1\)}%
    \mathinner{\mkern-2mu\left|{#1}\right\rangle\kern-0.3\ht\mstrut\rangle}%
}

\usepackage[colorlinks,linkcolor=blue,urlcolor=blue,citecolor=blue]{hyperref}

\begin{document}

\title{Optimal qubit-mediated quantum heat transfer via noncommuting operators and strong coupling effects}

\author{Marlon Brenes}
\email{marlon.brenes@ucr.ac.cr}
\affiliation{Department of Physics and Centre for Quantum Information and Quantum Control, University of Toronto, 60 Saint George St., Toronto, Ontario, M5S 1A7, Canada}
\affiliation{Centro de Investigación en Ciencia e Ingeniería de Materiales (CICIMA), Universidad de Costa Rica, San José, Costa Rica}
\affiliation{Escuela de Física, Universidad de Costa Rica, San José, Costa Rica}

\author{Jakub Garwo\l a}
\affiliation{Department of Physics and Centre for Quantum Information and Quantum Control, University of Toronto, 60 Saint George St., Toronto, Ontario, M5S 1A7, Canada}

\author{Dvira Segal}
\affiliation{Department of Chemistry, University of Toronto, 80 Saint George St., Toronto, Ontario, M5S 3H6, Canada}
\affiliation{Department of Physics and Centre for Quantum Information and Quantum Control, University of Toronto, 60 Saint George St., Toronto, Ontario, M5S 1A7, Canada}

\date{\today}

\begin{abstract}
Heat transfer in quantum systems is a current topic of interest due to emerging quantum technologies that attempt to miniaturize engines and examine fundamental aspects of thermodynamics. In this work, we consider heat transfer between two thermal reservoirs in which a central spin degree of freedom mediates the process. Our objective is to identify the system-bath coupling operators that maximize heat transfer at arbitrary system-bath coupling strengths. By employing a Markovian embedding method in the form of the reaction-coordinate mapping, we study numerically heat transfer at arbitrary system-bath coupling energy and for general system-bath coupling operators between the baths and the central qubit system. We find a stark contrast in the conditions required for optimal heat transfer depending on whether the system is weakly or strongly coupled to the heat baths. In the weak-coupling regime, optimal heat transfer requires identical coupling operators that facilitate maximum sequential transport, resonant with the central qubit. In contrast, in the strong-coupling regime, noncommuting system-bath coupling operators between the hot and cold reservoirs are necessary to achieve optimal heat transfer.  We further employ the Effective Hamiltonian theory and gain partial analytical insights into the observed phenomena. We discuss the limitations of this approximate method in capturing the behavior of the heat current for noncommuting coupling operators, calling for its future extensions to capture transport properties in systems with general interaction Hamiltonians.
\end{abstract}

\maketitle

\section{Introduction}
\label{sec:intro}

State-of-the-art technologies operating at the nanoscale attempt to employ exciting phenomena, such as the harnessing of quantum resources~\cite{Goold:2016}, to manipulate energy transfer in quantum thermal machines~\cite{Kosloff:2014Rev,Benenti:2017,Mitchison:2019}.
Conversely, thermal machines can be used to generate and manipulate quantum states \cite{Haack1, Haack2}.
In such configurations, one typically considers multiple thermal reservoirs held at different temperatures coupled to a central quantum system operating autonomously as a working medium. Beyond fundamental aspects at the level of our understanding of the thermodynamics, quantum thermal machines are experimentally realized in different platforms. In trapped-ion experiments~\cite{Rossnagel2016,Maslennikov2019,Lindenfels2019,VanHorne2020} the degree of control allows for single-atom working media. Nano-mechanical engines showed efficiencies beyond the standard Carnot limit---when coupled to squeezed thermal noise
~\cite{Klaers:2017}. Nitrogen vacancy centers in diamond~\cite{Maze:2008,BarGill12} also allow experimental testbeds for quantum thermal machines, in which quantum coherence plays a central role in thermal efficiency~\cite{Klatzow:2019}. More recently, demonstrations with noisy intermediate quantum computers allowed for a higher degree of control with respect to the working media in quantum thermal machines and their interactions with the bath~\cite{Chen:2024}. Simulating true thermal reservoirs still appears to be a challenge in such platforms, although progress has been made in this direction with current hardware~\cite{Motta2020}. Interestingly, superconducting qubits~\cite{Ronzani:2018,Senior:2020,Pekola24} and molecular junctions~\cite{PramodTE} also provide a platform for quantum heat engines with a focus on strong coupling effects. For example, in molecular junctions, the interactions between the molecule and the metal electrodes dictate the strength of the system-bath coupling. By controlling the chemical bonds at the contacts, it becomes possible to engineer strongly-coupled system-bath setups.

Fundamentally, a particular limit of interest is whenever the interaction energy of the coupling between the system and the baths is a small energy scale with respect to the system's energy scale and temperature, and the internal dynamics of the baths is fast relative to the system's characteristic relaxation timescale. In this case, the usual Born-Markov approximation ubiquitous in open systems theory~\cite{BreuerPetruccione} is justified. Assuming the system is coupled to two baths kept at different temperatures, in the limit of long times, the system reaches a nonequilibrium steady state (NESS). For central single-spin systems and bosonic reservoirs, it follows that in this weak coupling Markovian limit, the energy current in the NESS behaves as $j_{\rm NESS} \propto \lambda^2$, where $\lambda$ is a parameter that dictates the energy scale of the system-bath coupling configuration~\cite{Segal:2005,Segal:2006}. Furthermore, in this limit, the heat exchange processes are fully resonant, with energy quanta transferred between the baths in resonance with the central spin frequency. With respect to the central system, the consensus is that for single-atomic/qubit systems, it is required that the baths induce maximal sequential transport within the system such that optimal energy transfer can be achieved through these transitions~\cite{Kosloff:2013,Kosloff:2014Rev}. If the working medium, instead, corresponds to a many-body system; the transport properties of the entire system-bath configuration will typically depend on the microscopic details of the model. Heat transport may be coherent (ballistic) or incoherent (diffusive or anomalous) depending on the relevance of nonlinear interactions among each constituent of the many-body system~\cite{lepri2003thermal,dhar2008heat,chen2014nonintegrability,Brenes:2020TNT,barra2015thermodynamic,Znidaric:2010b,ZnidaricXXZspintransport}.

As the system-bath coupling parameter $\lambda$ is increased, the heat current increases due to the enhancement of the excitation/relaxation processes---until it turns over and becomes suppressed at higher values of $\lambda$.  This behavior, and other strong-coupling aspects of heat transport, have been demonstrated by developing and implementing a plethora of analytical and numerical techniques, including 
wavefunction based simulations \cite{Thoss08},
path integral methods~\cite{Kilgour:2019,Chen_2023}, Majorana fermion representations~\cite{Agarwalla:2017}, tensor network representations~\cite{Brenes:2020TNT,Popovic:2021}, Markovian embeddings~\cite{Nick:2021RC}, hierarchical equations of motion~\cite{Kato:2016,Kato2018,Ankerhold21,Petru24}, approximate analytical techniques~\cite{SBSegal1,SBSegal2},
mixed quantum-classical approaches \cite{Hanna1,Hanna2,Kelly},
quantum master equations~\cite{Wang:2015,Wang:2017,Grifoni24}, variational polaron transformations~\cite{Cao_2016,Hsieh:2019} and the Effective Hamiltonian  method~\cite{Segal:2023Effective,Nick:2023Effective}. At strong coupling, non-Markovian effects may be prominent, which hinder the applicability of Markovian theories. 

The scenario typically studied in the references mentioned before considers system-bath coupling operators that do not commute with the central system Hamiltonian. This condition is required for heat transfer~\cite{Kosloff:2013,Kosloff:2014Rev}. However, in most studies of quantum transport through impurities, the operators of the system that couple to the baths are equivalent for the cold ($\hat{S}^C$) and hot ($\hat{S}^H$) baths, such that (trivially) these operators {\it commute} with each other,  $[\hat{S}^C, \hat{S}^H] = 0$. A less explored direction corresponds to the case whenever $[\hat{S}^C, \hat{S}^H] \neq 0$, and particularly at a strong system-bath coupling energy. Previous studies have explored scenarios where the system-bath coupling operators are noncommuting~\cite{Duan:2020,NicknonC,Zhang:2024}, focusing on specific choices of $\hat{S}^C$ and $\hat{S}^H$. However, these works did not address the general configuration needed to determine the conditions for {\it optimal} heat transfer.

In this work, we consider quantum heat transfer across a junction in which the central system is composed of a single spin degree of freedom. We select a general class of system-bath coupling operators including cases in which $[\hat{S}^C, \hat{S}^H] \neq 0$, in order to investigate if different system-bath coupling operators have an effect on heat transfer at arbitrary system-bath coupling energy. Beyond the weak system-bath coupling energy limit, evaluating heat currents is usually a complicated task as nonresonant and non-Markovian effects may be prominent. To study the heat current at an arbitrary coupling energy, we adopt a general Markovian embedding method, known as the reaction coordinate mapping, which was applied for studies in, e.g., chemical reaction dynamics~\cite{Cao97,Hughes:2009RC,Hughes:2009RC2} and quantum thermodynamics \cite{Iles:2014RC,Strasberg:2016RC,Nick:2021RC}. This numerical approach has been shown to lead to correct behavior at arbitrary system-bath coupling energy, when tested for a class of models of the baths' spectral function~\cite{Iles:2014RC,Nick:2021RC,Nick:2023Effective,Brenes:2024Dynamics}. 

We observe a striking contrast in the behavior of the heat current depending on the energy scale of the system-bath coupling. At weak coupling, optimal heat transfer occurs when the system-bath coupling operators maximize sequential-resonant transport utilizing the energy transitions of the spin, achieved when the system is coupled to the two baths through specific identical operators. However, at arbitrary coupling strength, this is no longer the case. In fact, the coupling operators $\hat{S}$ that optimize heat transfer in weak couplings lead to {\it insulating} behavior in strong coupling. To maximize the heat current in the strong coupling regime, different coupling operators are required for hot and cold baths, satisfying $[\hat{S}^C, \hat{S}^H] \neq 0$. To address this behavior, we employ the recently developed Effective Hamiltonian theory for noncommuting system-bath coupling operators~\cite{Garwola:2024}. Through the dressing of operators, the method reveals the symmetry of specific configurations that promote optimal transport at strong coupling. As for the heat current, while the Effective method performs well for commuting operators, even in the strong-coupling regime, it fails to predict the correct behavior of the steady state heat current when $[\hat{S}^C, \hat{S}^H] \neq 0$. To understand this limitation, we present an analysis based on the eigenvalues of the relevant polaron-transformed system Hamiltonian. 

The rest of this article is organized as follows. Section~\ref{sec:transport} provides the fundamental description of the heat currents in the NESS for a generic open quantum system. It also includes a description of our numerical methodology to study strong-coupling thermodynamics, and results of simulations at weak and arbitrary strong system-bath coupling energies. Sec.~\ref{sec:effective} is devoted to the analytical Effective Hamiltonian theory for noncommuting system-bath coupling operators. We introduce the method and explain its predictions and limitations. We conclude with final remarks and future aspects in Sec.~\ref{sec:conc}.

\section{Heat transport with non-commuting coupling operators}
\label{sec:transport}

We consider the out-of-equilibrium dynamics of a quantum system $S$ coupled to two bosonic baths, each of which contains an infinite number of bosonic degrees of freedom. The total Hamiltonian is described by ($\hbar = 1$)
\begin{align}
\label{eq:h_total}
\hat{H} = \hat{H}_S &+ \hat{H}_I + \hat{H}_B \nonumber \\
=  \hat{H}_S &+ \sum_{\alpha=L,R} \hat{S}^{\alpha}\sum_k  t_{k,\alpha}  \left( \hat{c}_{k,\alpha}^{\dagger} + \hat{c}_{k,\alpha} \right) \nonumber \\
&+ \sum_{\alpha=L,R} \sum_k \nu_{k,\alpha} \hat{c}_{k,\alpha}^{\dagger} \hat{c}_{k,\alpha}.
\end{align}
We denote by $\alpha = L,R$ the {\em left} and {\em right} baths, respectively. Furthermore; $\hat{H}_S$ is the system Hamiltonian, the set $\{ \hat{c}_{k,n} \}$ are bosonic destruction operators for the $\alpha$-th bath, and $\nu_{k,\alpha}$ and $t_{k,\alpha}$ are the frequencies and the environment-system couplings between the $k$-th bosonic mode in the $\alpha$-th bath and the system, respectively. $\hat{S}^\alpha$ denotes the interaction operators, dubbed system-bath coupling operators, with support over the system's degrees of freedom. 
These system operators are coupled to the displacement operators of the baths. It is common to consider $\hat{S}^L = \hat{S}^R$; such that, trivially, $[\hat{S}^L, \hat{S}^R] = 0$. In this work, however, we investigate models in which $[\hat{S}^L, \hat{S}^R] \neq 0$.
Environmental effects are typically described via a spectral density function, which we define as \mbox{$J^{\alpha}(\omega) = \sum_k |t_{k,\alpha}|^2 \delta(\omega - \nu_{k,\alpha})$}.

To induce steady-state heat transfer, each bath is kept at different canonical equilibrium states,
\begin{align}
\hat{\rho}^{\alpha} = \frac{e^{-\beta_{\alpha} \hat{H}_{B,\alpha}}}{Z_\alpha},
\end{align}
where $\beta_{\alpha} = 1 / T_{\alpha}$ is the inverse temperature of the $\alpha$-th bath, $\hat{H}_{B,\alpha} = \sum_k \nu_{k,\alpha} \hat{c}_{k,\alpha}^{\dagger} \hat{c}_{k,\alpha}$ and $Z_\alpha = \Tr[e^{-\beta_{\alpha} \hat{H}_{B,\alpha}}]$ is the partition function for the $\alpha$-th bath. 

In Sec.~\ref{sec:ness}, we focus on the weak coupling approach to quantum heat transport. In Sec.~\ref{sec:strong} we target the general coupling regime. To capture strong-coupling processes, we introduce the reaction coordinate mapping and the resulting Markovian embedding tool. Sec.~\ref{sec:res} presents results spanning the weak- to strong-coupling regimes.

\subsection{Non-equilibrium steady states and heat currents}
\label{sec:ness}

The out-of-equilibrium configuration imposed on the baths induces a {\em nonequilibrium steady state} onto the system. These states occur generically whenever the system Hamiltonian is time-independent. Such states may be understood from the dynamics of the entire system-bath configuration. 

Under the Born-Markov approximations, a master equation of the Redfield type may be obtained for the dynamics of the reduced state of the system $\hat{\rho}(t)$. The equation is given by~\cite{NitzanBook}
\begin{align}
\label{eq:redfield_dyn}
\frac{{\rm d}\rho_{mn}}{{\rm d}t} &= -\ii \omega_{mn} \rho_{mn} \nonumber \\
&- \sum_{\alpha jk} \left[ R^{\alpha}_{mj,jk}(\omega_{kj}) \rho_{kn} + [R^{\alpha}_{nk,kj}(\omega_{jk})]^* \rho_{mj} \right. \nonumber \\
&- \left. R^{\alpha}_{kn,mj}(\omega_{jm}) \rho_{jk} - [R^{\alpha}_{jm,nk}(\omega_{kn})]^* \rho_{jk} \right],
\end{align} 
where $\omega_m$ are eigenenergies of the system Hamiltonian, from $\hat{H}_S \ket{\omega_m} = \omega_m \ket{\omega_m}$. In Eq.~\eqref{eq:redfield_dyn}, the subindices refer to the matrix elements in the energy eigenbasis of $\hat{H}_S$, while $R_{ab,cd}$ are the elements in this basis of the Redfield tensor. In the above expression, we have defined $\omega_{mn} = \omega_m - \omega_n$. In deriving the Redfield equation, one relies on the assumption that the initial state of the total system+baths configuration at time $t = t_0$ is a tensor product of the system and each individual bath; i.e., there are no system-bath or bath-bath correlations at time $t = t_0$~\cite{BreuerPetruccione}.  

The Redfield tensor is evaluated from the bath equilibrium correlation functions and, crucially, it depends on the system-bath coupling operators via
\begin{align}
\label{eq:refield_tensor_integral}
R^{\alpha}_{mn,jk}(\omega) &= S^{\alpha}_{mn} S^{\alpha}_{jk} \int_{0}^{\infty} {\rm d}\tau e^{\ii \omega \tau} \langle \hat{B}^{\alpha}(\tau) \hat{B}^{\alpha}(0) \rangle,
\end{align}
where $\hat{B}^{\alpha}(\tau) = \sum_k  t_{k,\alpha}  \left( \hat{c}_{k,\alpha}^{\dagger} e^{\ii \nu_{k,\alpha} \tau} + \hat{c}_{k,\alpha} e^{-\ii \nu_{k,\alpha} \tau} \right)$ corresponds to the interaction operator with support over the bath degrees of freedom. We note that we have assumed that the baths are stationary, with the condition $[\hat{\rho}^{\alpha}, \hat{H}_{B,\alpha}] = 0$, such that the elements of the Redfield tensor do not depend explicitly on time~\cite{BreuerPetruccione}.

In the continuum limit, we have 
\begin{align}
\label{eq:refield_tensor_correlation}
R^{\alpha}_{mn,jk}(\omega) &= S^{\alpha}_{mn} S^{\alpha}_{jk} \left[ \Gamma^{\alpha}(\omega) + \ii \delta^{\alpha}(\omega) \right],
\end{align}
where $\Gamma^{\alpha}(\omega)$ is the symmetric part of the correlation function and $\delta^{\alpha}(\omega)$ is the Lamb shift term. We neglect the latter term in our calculations~\cite{Correa_Lamb}. In order to evaluate $\Gamma^{\alpha}(\omega)$, one needs to consider a spectral function. In our calculations, we look at a Brownian spectral density
\begin{align}
\label{eq:spec_fun_brownian}
J^{\alpha}(\omega) = \frac{4\gamma_{\alpha} \Omega_{\alpha}^2 \lambda_{\alpha}^2 \omega}{(\omega^2 - \Omega_{\alpha}^2)^2 + (2\pi\gamma_{\alpha}\Omega_{\alpha}\omega)^2},
\end{align}
which is peaked at $\Omega_{\alpha}$ with a width parameter $\gamma_{\alpha}$. In this notation, $\lambda_{\alpha}$ corresponds to the system-bath coupling to the $\alpha$-th bath. In Sec.~\ref{sec:strong}, we shall see that this spectral function is amenable to study strong-coupling thermodynamics.

Equation~\eqref{eq:redfield_dyn} describes the entire dynamics of the reduced state of the system $\hat{\rho}(t)$, however, in this work we are interested in the long-time limit. In this limit, the state becomes independent of time and gives rise to the nonequilibrium steady state
\begin{align}
\lim_{t \to \infty} \hat{\rho}(t) = \hat{\rho}_{\rm NESS}.
\end{align}
In a more succinct representation, we may express Eq.~\eqref{eq:redfield_dyn} as
\begin{align}
\frac{{\rm d}\hat{\rho}(t)}{{\rm d}t} = -\ii [\hat{H}_S, \hat{\rho}(t)] + \mathcal{R}^L\{ \hat{\rho}(t) \} + \mathcal{R}^R\{ \hat{\rho}(t) \},
\end{align}
where we have split the contributions to the dynamics of the system into a coherent part and the dissipative effect of both left and right baths, by the actions of the superoperators $\mathcal{R}^L$ and $\mathcal{R}^R$. Note that this is an equivalent representation to the one shown in Eq.~\eqref{eq:redfield_dyn}. However, this representation allows us to evaluate the heat currents. In the NESS, we have
\begin{align}
\label{eq:heat}
j^{\alpha}_{\rm NESS} = \Tr[\mathcal{R}^{\alpha}\{\hat{\rho}_{\rm NESS}\} \hat{H}_S].
\end{align}
Without loss of generality, we shall consider the cases in which $T_L > T_R$ such that the heat flows from the left bath to the right bath. The NESS is characterized by $j^L_{\rm NESS} = -j^R_{\rm NESS} = j_{\rm NESS}$. 

It is then left to define how to evaluate the NESS. One could evaluate the dynamics of the reduced state of the system generated by Eq.~\eqref{eq:redfield_dyn} up to times long enough up until the condition $j^L_{\rm NESS} = -j^R_{\rm NESS}$ is satisfied. However, there is a better procedure that yields the NESS with lower computational cost. With the eigendecomposition in Eq.~\eqref{eq:redfield_dyn}, one can express the reduced density matrix of the state $\hat{\rho}(t)$ as a vector by concatenating the columns of the density matrix, with $\mathcal{R}$ a corresponding matrix that acts as an operator in the vectorized representation. It follows that the Redfield equation may be written as
\begin{align}
\frac{{\rm d}  \kett{\hat{\rho}(t)}}{{\rm d}t} = \hat{L} \kett{\hat{\rho}(t)},
\end{align}
where $\kett{\hat{\rho}(t)}$ is the vectorized density matrix and $\hat{L}$ is the matrix representation of the superoperator that contains the coherent and incoherent contributions from internal and bath degrees of freedom. In the NESS, we have 
\begin{align}
\hat{L} \kett{\hat{\rho}_{\rm NESS}} = 0, 
\end{align}
since the state is stationary in time. We can obtain this solution by solving a system of linear equations constrained to the trace-preserving property of the density matrix $\bbra{\mathds{1}} \kett{\hat{\rho}} = \Tr[\hat{\rho}] = 1$, where $\kett{\mathds{1}}$ is the vectorized identity. One can then define $\tilde{L} = \hat{L} + \kett{0} \bbra{\mathds{1}}$ such that
\begin{align}
\label{eq:vectorised}
&\tilde{L}\kett{\hat{\rho}_{\rm NESS}} = \hat{L}\kett{\hat{\rho}_{\rm NESS}} + \kett{0} \bbra{\mathds{1}}\kett{\hat{\rho}_{\rm NESS}} \nonumber \\
\implies &\tilde{L}\kett{\hat{\rho}_{\rm NESS}} = \kett{0},
\end{align}
where $\kett{0}$ is the vectorized form of the first state in the vector space. This choice is arbitrary and any other state would suffice. Eq.~\eqref{eq:vectorised} can then be readily solved by standard techniques such as $LU$ decomposition followed by back-substitution to obtain the NESS.

\subsection{Strong coupling}
\label{sec:strong}

The procedure described in Sec.~\ref{sec:ness} rests on the assumption that the coupling energy ($\lambda_{\alpha}$, in our notation) is the smallest energy scale in the problem, such that the Born-Markov assumptions can be justified and the standard open systems theory may be applied~\cite{BreuerPetruccione}. At strong system-bath coupling energy, these assumptions cannot be justified.

We may, however, study strong-coupling thermodynamics including non-Markovian effects via a \emph{Markovian embedding} theory. Within this framework, some degrees of freedom of the bath are explicitly included as part of the system. The resulting system+baths configuration after this procedure is carried out can be investigated under the Born-Markov approximation, so long as the coupling between the \emph{extended} system and the \emph{residual} baths now becomes a perturbative energy scale. There exist different formalisms that are based on this procedure, one of which corresponds to the \emph{reaction coordinate} (RC) \emph{mapping}.

The RC mapping was originally developed to handle quantum effects in electron reaction rates under strong (electron-nuclei) couplings (see Refs. \cite{Cao97,Hughes:2009RC,Hughes:2009RC2} for some examples).
The process of mapping
starts by extracting a single mode from each bath and including it as part of the system. After the mapping, Eq.~\eqref{eq:h_total} translates to~\cite{Cao97, Hughes:2009RC,Hughes:2009RC2,Martinazzo:2011RC,Iles:2014RC,NazirJCP16,Strasberg:2016RC,Nick:2021RC,Segal:2023Effective,Luis19,Gernot19}
\begin{align}
\label{eq:h_rc_tot}
&\hat{H}^{\rm{RC}} = \hat{H}_S + \sum_{\alpha} \left[ \Omega_{\alpha} \hat{a}^{\dagger}_{\alpha} \hat{a}_{\alpha} + \lambda_{\alpha} \hat{S}^{\alpha} \left( \hat{a}^{\dagger}_{\alpha} + \hat{a}_{\alpha} \right) \right] \nonumber \\
&+ \sum_{\alpha k} f_{k,\alpha} \left( \hat{a}^{\dagger}_{\alpha} + \hat{a}_{\alpha} \right) \left( \hat{b}_{k,\alpha}^{\dagger} + \hat{b}_{k,\alpha} \right) + \sum_{\alpha k} \omega_{k,\alpha} \hat{b}_{k,\alpha}^{\dagger} \hat{b}_{k,\alpha}.
\end{align}
The bosonic operators $\hat{a}_{\alpha}$ and $\hat{a}^{\dagger}_{\alpha}$ refer to the reaction coordinates of the $\alpha$-th bath with frequencies $\Omega_{\alpha}$; while $\lambda_{\alpha}$ is the coupling between the original system $\hat{H}_S$ and both reaction coordinates. We recall that $\hat{S}^{\alpha}$ refers to the system's coupling operator to the baths, however, in this representation, the coupling occurs directly to the reaction coordinates. The so-called \emph{enlarged} system is now composed of $\hat{H}_S$, the reaction coordinates, and their coupling,
\begin{align}
\label{eq:h_rc_s}
\hat{H}^{\rm{RC}}_S = \hat{H}_S + \sum_{\alpha} \left[ \Omega_{\alpha} \hat{a}^{\dagger}_{\alpha} \hat{a}_{\alpha} + \lambda_{\alpha} \hat{S}^{\alpha} \left( \hat{a}^{\dagger}_{\alpha} + \hat{a}_{\alpha} \right) \right].
\end{align} 
The remaining terms in Eq.~\eqref{eq:h_rc_tot} correspond to the \emph{residual} baths and the coupling between these baths and the reaction coordinates. The coupling is now represented by a different spectral function $J^{\alpha}_{\rm RC}(\omega) = \sum_k |f_{k,\alpha}|^2 \delta(\omega - \omega_k)$. The harmonic modes of the residual bath are represented via the canonical operators $\{ \hat{b}_k \}$ with frequencies $\omega_k$. One can derive both $\lambda_{\alpha}$ and $\Omega_{\alpha}$ from the original spectral function via $\lambda^2_{\alpha} = \frac{1}{\Omega}_{\alpha} \int_0^\infty {\rm{d}}\omega \; \omega J^{\alpha}(\omega)$ and $\Omega^2_{\alpha} = \frac{\int_0^\infty {\rm{d}}\omega \; \omega^3 J^{\alpha}(\omega)}{\int_0^\infty {\rm{d}}\omega \; \omega J^{\alpha}(\omega)}$~\cite{Iles:2014RC}. 
The mapping is general and can be used for different spectral functions. However, for our choice of Brownian spectral function in Eq.~\eqref{eq:spec_fun_brownian}, the resulting spectral function after the mapping has been shown to be Ohmic~\cite{Iles:2014RC},
\begin{align}
\label{eq:spec_fun_rc_ohmic}
J^{\alpha}_{\rm RC}(\omega) = \gamma_{\alpha} \omega e^{-\omega / \Lambda_{\alpha}},
\end{align}
where $\Lambda_{\alpha}$ is a large energy cut-off and $\gamma_{\alpha}$ is a dimensionless parameter representing the coupling to the $\alpha$-th residual bath. 
Importantly, the spectral density function after the reaction coordinate mapping does not depend on the original coupling parameter $\lambda$. As such, enhancing this interaction increases terms within the extended-system Hamiltonian, but not between the extended system and the residual bath.

With the spectral function of the mapped model, $\Gamma^{\alpha}(\omega)$ can be evaluated directly to obtain
\begin{align}
\Gamma_{\rm RC}^{\alpha}(\omega) = \begin{cases} \pi J^{\alpha}_{\rm RC}(|\omega|) n(|\omega|) & \omega < 0 \\ \pi J^{\alpha}_{\rm RC}(\omega) [n(\omega) + 1] & \omega > 0 \\ \pi \gamma_{\alpha} / \beta & \omega = 0, \end{cases}
\end{align}
where we have defined $n(\omega) = (e^{\beta \omega} - 1)^{-1}$ as the Bose-Einstein distribution function. 

Under the assumption that the coupling of the extended system to the residual bath is weak,
one employs the same Redfield Eq.~\eqref{eq:redfield_dyn} to study strong coupling thermodynamics, with an arbitrary large $\lambda$. In fact, all the steps described in Sec.~\ref{sec:ness} follow at strong coupling via this Markovian embedding. The differences to be noted correspond to the extended system Hamiltonian Eq.~\eqref{eq:h_rc_s}, which comprises both the original system and the reaction coordinates. Furthermore, the interaction Hamiltonian of the residual baths is now given by $\hat{S}^{\alpha}_{\rm res} = \hat{a}^{\dagger}_{\alpha} + \hat{a}_{\alpha}$, which is employed to evaluate the Redfield tensor in Eq.~\eqref{eq:refield_tensor_correlation}. 

We note that, in the absence of a general analytical solution to Eq.~\eqref{eq:redfield_dyn} for generic system-bath coupling operators and system Hamiltonian, we approach the problem numerically. Specifically, using simulations, we solve Eq.~\eqref{eq:vectorised} to obtain the NESS and then Eq.~\eqref{eq:heat} to evaluate the heat currents. This numerical approach requires truncating the number of reaction-coordinate levels to a finite value, $M$.

The reaction coordinate mapping approach described here can be extended in several ways, making it more general and more accurate.
The idea of chain mapping is the most natural extension. In this approach, the RC transformation is performed iteratively, each round extracting an additional harmonic degree of freedom from the bath \cite{Woods:2014Embedding,chainM1, chainM2}. This would allow for accurately capturing structured spectral density functions.

\begin{figure}[t]
\fontsize{13}{10}\selectfont 
\centering
\includegraphics[width=\columnwidth]{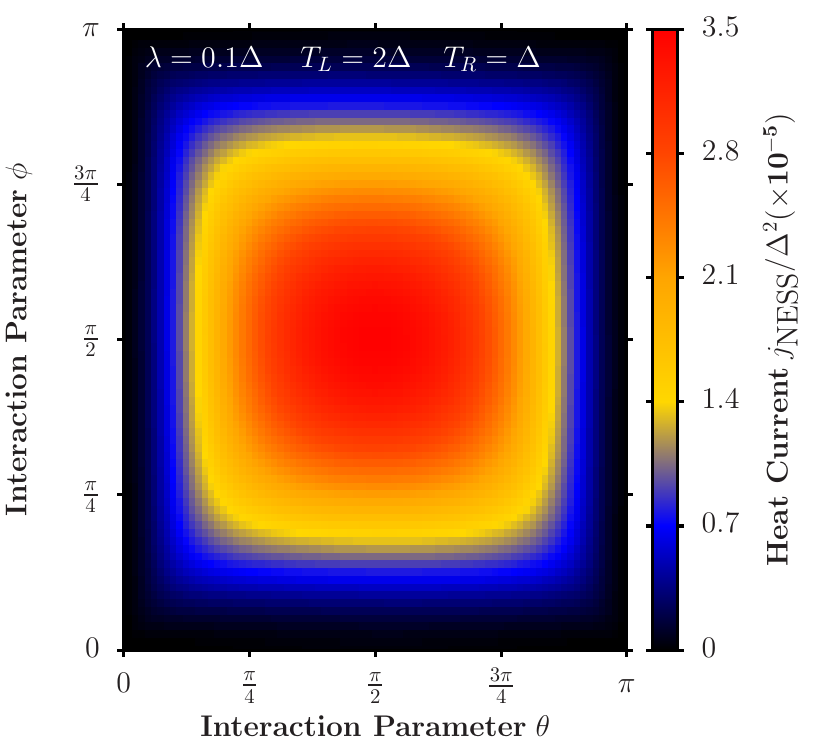}
\caption{Heat current in the NESS flowing from left ($T_L = 2\Delta$) towards right bath ($T_R = \Delta$) at weak system-bath coupling $\lambda_L = \lambda_R = \lambda = 0.1\Delta$ as a function of the interaction parameters $(\theta, \phi)$. For this calculation we used $\Omega_L = \Omega_R = \Omega = 8\Delta$, $M = 6$ and $\gamma_L = \gamma_R = 0.05 / \pi$.}
\label{fig:1}
\end{figure}

\subsection{Results}
\label{sec:res}

\subsubsection{Heat transfer at weak coupling}
\label{sec:res_weak}

\begin{figure*}[t]
\fontsize{13}{10}\selectfont 
\centering
\includegraphics[width=0.32\textwidth]{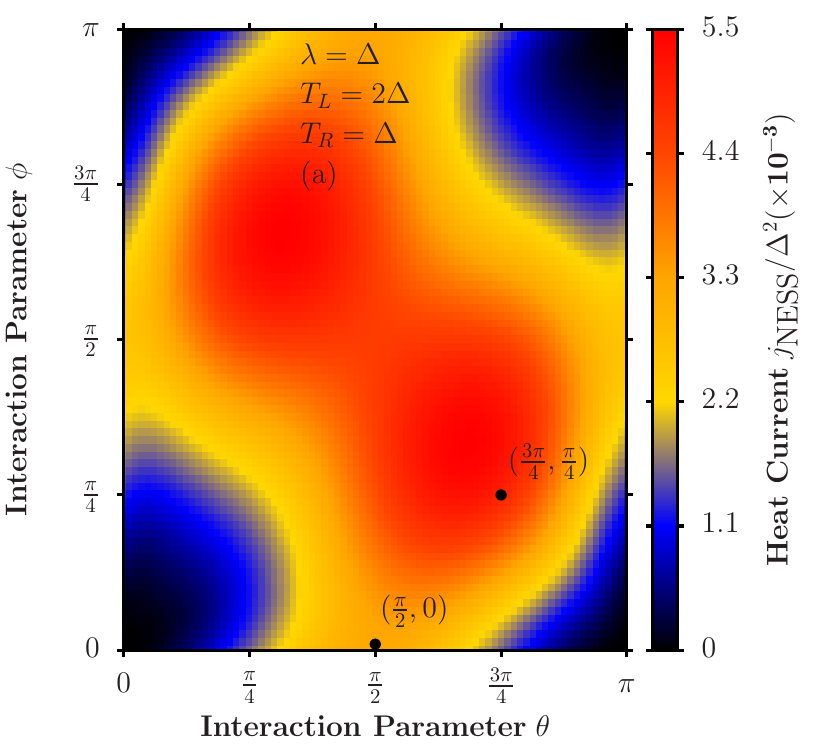}
\includegraphics[width=0.32\textwidth]{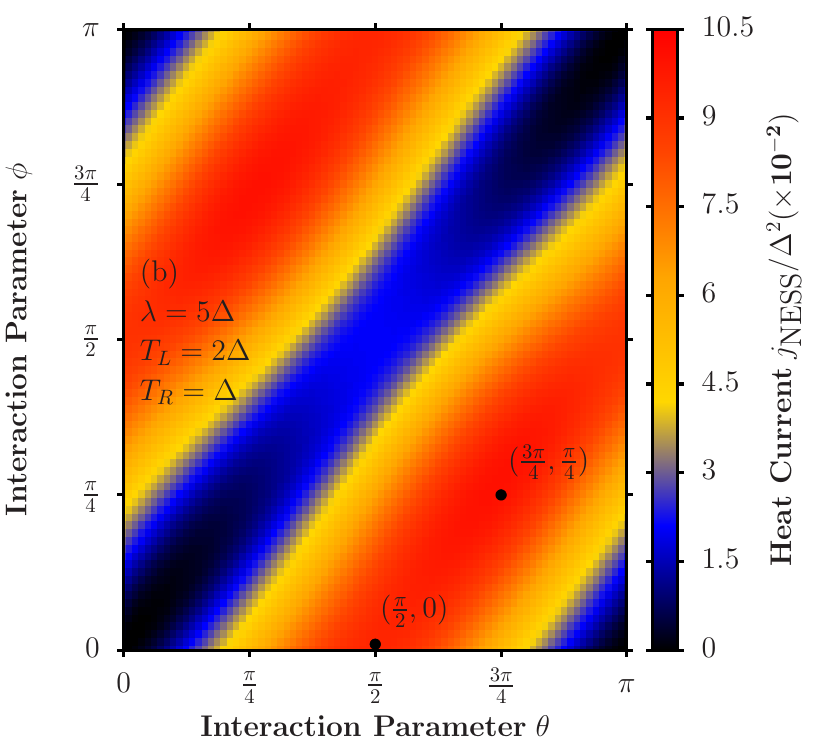}
\includegraphics[width=0.32\textwidth]{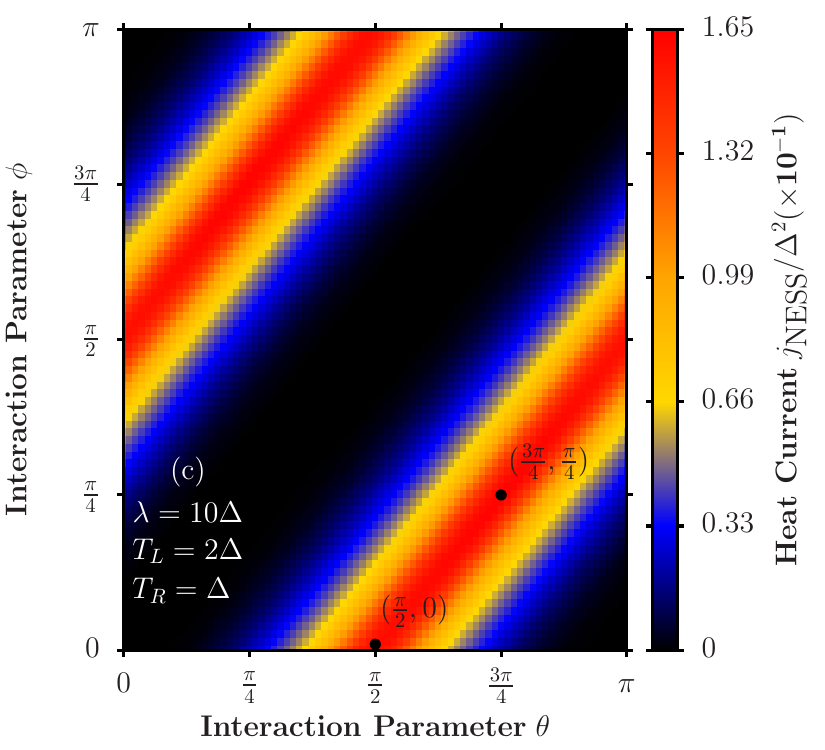}
\caption{Heat current in the NESS flowing from left ($T_L = 2\Delta$) towards right bath ($T_R = \Delta$) at strong system-bath coupling (a) $\lambda_L = \lambda_R = \lambda = \Delta$, (b) $\lambda = 5\Delta$ and (c) $\lambda = 10\Delta$ as a function of the interaction parameters $(\theta, \phi)$. For this calculation we used $\Omega_L = \Omega_R = \Omega = 8\Delta$, $M = 6$ and $\gamma_L = \gamma_R = 0.05 / \pi$ (note that the values of the current have been rescaled by a different constant value in each panel).}
\label{fig:2}
\end{figure*}

We now turn to investigating the NESS heat currents evaluated from Eq.~\eqref{eq:vectorised} and Eq.~\eqref{eq:heat} in our dual-bath configuration. We are interested in the general case such that $[\hat{S}^L, \hat{S}^R] \neq 0$. Our central system is a single spin degree of freedom
\begin{align}
\hat{H}_S = \Delta \hat{\sigma}^z,
\end{align}
where $\Delta$ denotes the spin-splitting energy. In order to consider a general set of noncommuting system-bath coupling operators, we define
\begin{align}
\label{eq:coupling_angles}
\hat{S}^L &= \cos(\theta) \hat{\sigma}^z + \sin(\theta) \hat{\sigma}^x, \nonumber \\
\hat{S}^R &= \cos(\phi) \hat{\sigma}^z + \sin(\phi) \hat{\sigma}^x,
\end{align}  
such that we can control noncommutativity between the $\hat S_{L,R}$ operators via two angle parameters $(\theta, \phi)$. 

At weak coupling, transport takes place through the eigenstates of the system Hamiltonian; which in this case are $\ket{\uparrow}$ and $\ket{\downarrow}$. Effectively, energy transfer is induced by the transitions between the energy levels of the system, and transport is sequential and resonant with the qubit splitting $2\Delta$, see discussion in Appendix \ref{ap:2}. 

In Fig.~\ref{fig:1}, we present simulations of the heat current in the weak coupling regime, displayed as a function of the interaction angles $(\theta, \phi)$. The current at weak coupling is at its peak whenever $(\theta, \phi) = \left(\frac{\pi}{2}, \frac{\pi}{2}\right)$, i.e., 
when we have $\hat{S}^L = \hat{S}^R = \hat{\sigma}^x$. In this case, the hot bath excites the system, and the qubit medium transfers these excitations to the cold bath. Moving away from the point $(\theta, \phi) = \left(\frac{\pi}{2}, \frac{\pi}{2}\right)$ only leads to sub-optimal versions of this process. This is because the contribution to the sequential process weakens and the inclusion of $\hat{\sigma}^z$ terms does not contribute to the excitation current that moves through the system.

\subsubsection{Heat transfer at strong coupling: the role of the coupling operators}
\label{sec:res_strong}

For simplicity, we assume the coupling energy parameter is equal at the two sides, $\lambda = \lambda_L = \lambda_R$. At weak coupling, we found that the system operator that couples to the baths should be kept close to $\hat{S}^L = \hat{S}^R = \hat{\sigma}^x$ for maximum sequential transport. This, in general, is supported by the fact that heat transfer is mediated by the transitions generated within the system, such that the choice $\hat{S}^L = \hat{S}^R = \hat{\sigma}^x$ is justified.

In Fig.~\ref{fig:2} we show the simulated heat current for the same junction's parameters as in Fig~\ref{fig:1}. In this case, however, we increase the coupling energy $\lambda$ to understand strong-coupling behavior. In Fig.~\ref{fig:2}(a) we see that increasing the coupling energy for the selected coupling operators in Eq.~(\ref{eq:coupling_angles}) means that in order to attain maximum energy transfer one needs to consider different operators other than $\hat{S}^L = \hat{S}^R = \hat{\sigma}^x$. In fact, at stronger values of $\lambda$ as shown in Fig.~\ref{fig:2}(b) and Fig.~\ref{fig:2}(c), the maximum value of the current is continuously attained closer to the points for which the interaction parameters are out of phase by a value of $\pi / 2$; that is,
\begin{align}
\label{eq:angles_dephased_pi}
\hat{S}^L &= \cos(\delta) \hat{\sigma}^z + \sin(\delta) \hat{\sigma}^x, \nonumber \\
\hat{S}^R &= \cos\left(\delta \pm \frac{\pi}{2} \right) \hat{\sigma}^z + \sin\left(\delta \pm \frac{\pi}{2}\right) \hat{\sigma}^x \nonumber \\
&= \mp \sin(\delta) \hat{\sigma}^z \pm \cos(\delta) \hat{\sigma}^x,
\end{align}
where $\delta \in [0, \pi]$ is an arbitrary angle chosen to control the interaction operator. 

Although the heat current in Fig.~\ref{fig:2} appears to be symmetric under the inversion of the coupling operators $\hat{S}^L$ and $\hat{S}^R$, a careful comparison shows that this is not the case. Stated differently, we observe a difference in the magnitude of the heat current when the temperatures are exchanged $T_L \to T_R$ and $T_R \to T_L$, as long as the coupling operators are different, and even when $\lambda_L=\lambda_R$.
Notably, this difference becomes more pronounced as the temperature bias increases, indicating that the system functions as a heat current rectifier. This observation agrees with the findings reported in Ref.~\cite{Duan:2020}.

These results reveal that noncommuting interaction operators are relevant for maximum heat transfer at strong system-bath coupling energy. At even stronger values of $\lambda$, the heat current resembles the results obtained for $\lambda = 10\Delta$ from Fig.~\ref{fig:2}(right-most panel); with the exception of a narrower maximum along the $(\delta, \delta \pm \frac{\pi}{2})$ points, see Appendix~\ref{ap:1} for further results. We note that in Fig.~\ref{fig:2} the absolute values of the energy current are $2-4$ orders of magnitude higher than the weak coupling results in Fig.~\ref{fig:1}.

We conclude this part with some physical insight into our observations.
When the system is coupled to both baths via its $\hat\sigma_x$ operator, the heat current ultimately vanishes at strong coupling, as has been reported using various methods (e.g., Ref. \cite{Nick:2021RC} and references therein). 
One way to interpret this effect is that, under strong coupling, the spin splitting is effectively completely suppressed, and thus heat cannot be exchanged between the system and the baths. Another point of view relies on the picture of polarons: At ultrastrong coupling, the excitation of the two-level system, which is dressed by the phonons of the baths, becomes exceedingly heavy, such that it is effectively localized, inhibiting transport.

In contrast, when the couplings at the two baths are orthogonal,  through $\hat\sigma_z$ and $\hat\sigma_x$, heat cannot transfer sequentially between the qubit system and the $\hat\sigma_z$-coupled contact. Rather, the system serves to mediate energy transfer between the baths, with the current scaling as $\lambda^4$ (at weak coupling); for details see Ref. \cite{NicknonC}. While we do not have concrete scaling relations for the current at ultrastrong coupling,  bath-to-bath transport is still the only transport route in the ``orthogonal" $(\theta,\phi)$=$(\delta,\delta\pm\pi/2)$ case.
\subsubsection{Dependence on the coupling energy}
\label{sec:res_lambda_dependence}

We found a stark contrast in the behavior of the overall heat transfer between the weak- and strong-coupling regimes. In the weak-coupling regime, optimal transport requires coupling operators that facilitate transitions between the energy levels of the system. In contrast, in the strong-coupling regime, maximum heat transfer is achieved with coupling operators that do not commute.

\begin{figure}[b]
\fontsize{13}{10}\selectfont 
\centering
\includegraphics[width=\columnwidth]{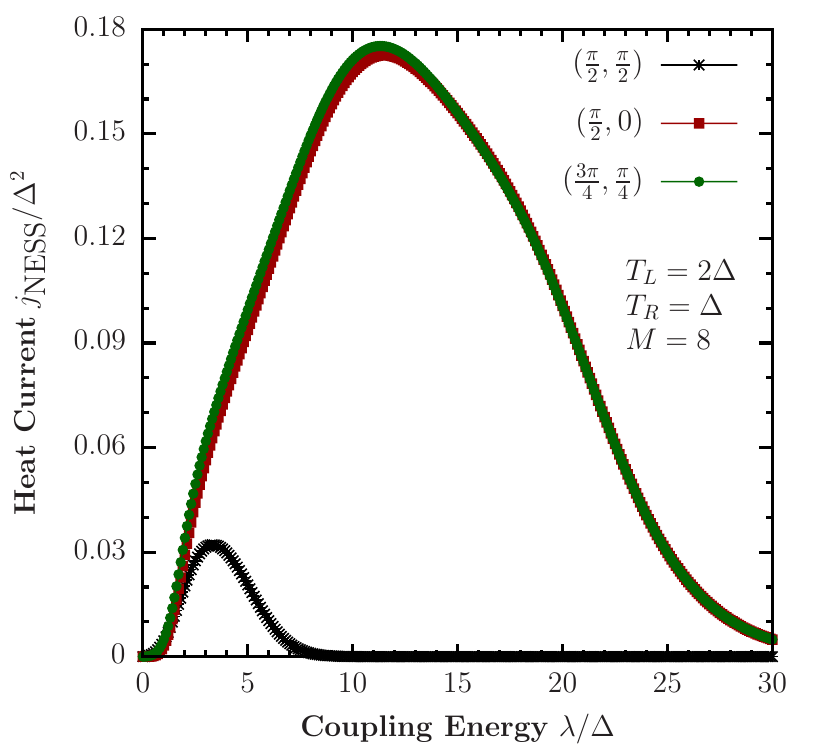}
\caption{Heat current in the NESS flowing from left ($T_L = 2\Delta$) towards right bath ($T_R = \Delta$) as a function of system-bath coupling $\lambda_L = \lambda_R = \lambda$. Both interaction angles $(\theta, \phi) = (\frac{\pi}{2}, 0)$ and $(\theta, \phi) = (\frac{3\pi}{4}, \frac{\pi}{2})$, satisfy the condition of Eq.~\eqref{eq:angles_dephased_pi} for which the heat current attains its maximum value at strong system-bath coupling, see Fig.~\ref{fig:2}. We also present the commuting model with $(\theta, \phi) = (\frac{\pi}{2}, \frac{\pi}{2})$, reaching here a lower maximal current at smaller $\lambda$, compared to the other two non-commuting models. For this calculation we used $\Omega_L = \Omega_R = \Omega = 8\Delta$, $M = 8$ and $\gamma_L = \gamma_R = 0.05 / \pi$.}
\label{fig:3}
\end{figure}

Following our findings from Sec.~\ref{sec:res_strong}, we now investigate the behavior of the heat current as a function of the coupling energy parameter $\lambda_L = \lambda_R = \lambda$. In Fig.~\ref{fig:3}, we present the heat current as a function of $\lambda$ for three different choices of interaction angles. Let us begin with the noncommuting models, $(\theta, \phi) = (\frac{\pi}{2}, 0)$ and $(\theta, \phi) = (\frac{3\pi}{4}, \frac{\pi}{2})$. In both of these cases, the heat current attains its maximum value (in angle space), see results in Fig.~\ref{fig:2}(b) and Fig.~\ref{fig:2}(c). As a function of $\lambda$, the heat current displays a nonmonotonic behavior, initially increasing as $\lambda$ increases, then turning over at sufficiently strong $\lambda$. In Fig.~\ref{fig:3}, we also display the heat current as a function of $\lambda$ for the commuting operator case $(\theta, \phi) = (\frac{\pi}{2}, \frac{\pi}{2})$, for which $\hat{S}^L = \hat{S}^R = \hat{\sigma}^x$. We recall that at weak coupling, these are the coupling operators that lead to optimal heat transfer (see Fig.~\ref{fig:1}). At intermediate-to-strong coupling, however, we observe that the maximum heat current is an order of magnitude higher if we introduce noncommuting operators on the left and right baths, rather than the more commonly-considered case of $\hat{S}^L = \hat{S}^R = \hat{\sigma}^x$. Furthermore, it is interesting to note that the value of the system-bath coupling $\lambda$ for which the maximum heat current is reached differs (shifts) depending on the system-bath coupling operators chosen, commuting or noncommuting. 
In the next section, we employ an analytical framework that helps shedding light on this behavior.

In Appendix~\ref{ap:3}, we exemplify convergence trends in reaction coordinate simulations for commuting and noncommuting models with respect to $M$, the number of levels maintained in the reaction coordinate harmonic manifold. Not only are the currents distinct in these cases, but also their numerical convergence is markedly different. This will be rationalized in the next section as well; see Fig.~\ref{fig:6}.

We now discuss a potential experimental setup to observe
the tunability of the heat current with coupling operators (Figs. \ref{fig:1} and \ref{fig:2}). The reaction coordinate Hamiltonian, Eq. (\ref{eq:h_rc_tot}), can be naturally implemented using a superconducting qubit coupled to two resonators, each in contact with its own thermal reservoir. This setup was realized in Ref. \cite{Pekola24}, where strong coupling between a flux qubit and its attached resonators was achieved, leading to thermal transport far exceeding that observed in the weak-coupling regime~\cite{Senior:2020}. In the experiment, a coupling ratio of  $\lambda/\Delta\approx 0.1$ and temperature $T\approx \Delta$, were attained, 
with temperature difference $|T_L-T_R|  \approx \Delta$,
conditions similar to those used here. Under such conditions, one should be able to observe the nontrivial behavior of heat flux as a function of the coupling operators, as illustrated in Figs. \ref{fig:1} and Fig. \ref{fig:2}(a).

However, two important considerations arise when comparing our theory-simulation study to existing experimental devices~\cite{Senior:2020,Pekola24}. First, in simulations, we adopted high-frequency resonators ($\Omega>\Delta$), whereas the resonator frequencies in Ref. \cite{Pekola24} were of the same order as the qubit's gap, $\Delta$. This distinction is likely not critical,  and simulations could be adapted to more closely match the experimental regime. The second and more significant challenge concerns the nature of the interaction operators. Our work here identifies specific combinations of dispersive–dissipative  ($\hat\sigma_z$-$\hat\sigma_x$) coupling operators as optimal for maximizing thermal transport. Although experimental coupling of a superconducting qubit to two baths through noncommuting operators has been demonstrated in the context of quantum computing \cite{Kamal24}, realizing tunable coupling operators of the qubit to the resonators remains a technical challenge.
 
\section{Effective Hamiltonian Theory}
\label{sec:effective}

The transport behavior at weak coupling stands in stark contrast to that observed at strong coupling. In the weak-coupling regime, optimal heat transfer requires system-bath coupling operators that facilitate transitions between the energy levels of a two-level system.
This can be achieved with $\hat{S}^L = \hat{S}^R = \hat{\sigma}^x$ for our system Hamiltonian $\hat{H}_S = \Delta \hat{\sigma}^z$ (see Fig.~\ref{fig:1}). Interestingly, this is no longer the case whenever the system-bath coupling energy $\lambda$ increases arbitrarily. We find that, instead, one requires operators that do not commute with each other, i.e., $[\hat{S}^L, \hat{S}^R] \neq 0$. Even more interesting is that in our configuration there appears to exist a symmetry, such that if one defines $\hat{S}^L$ and $\hat{S}^R$ from Eq.~\eqref{eq:coupling_angles}, the optimal heat transfer occurs whenever the interaction parameters are out of phase by an angle of $\pi / 2$ as in Eq.~\eqref{eq:angles_dephased_pi}, see Fig.~\ref{fig:2}. 
In what follows, we analyze these findings from the perspective of the Effective Hamiltonian theory.

\subsection{Effective system and system-bath operators}
\label{sec:outline_effective}

The Effective Hamiltonian method, first introduced in Ref.~\cite{Segal:2023Effective}, was developed with the goal of providing analytical insights into strong coupling effects.

It was previously applied to study dynamical, equilibrium and nonequilibrium steady-state problems~\cite{Segal:2023Effective,Nick:2023Effective,Brenes:2024Dynamics,Min:2024Phases,Garwola:2024}, showing good success in providing qualitative and even quantitative-correct results: For thermal transport problems, the Effective Hamiltonian method was used in Ref.~\cite{Segal:2023Effective}, but only in the case of commuting system-bath operators, where it provided results in excellent agreement with other established numerical simulations~\cite{Nick:2021RC}. 
Building on the recent extension of the Effective Hamiltonian method to handle multiple noncommuting operators~\cite{Garwola:2024}, we apply it to the current problem in order to provide physical insights. 

We introduce the Effective Hamiltonian theory in this section in a brief outline with the most important results. For further details, we refer the reader to Appendix~\ref{ap:2}. For an extensive description and further results, we refer the reader to Refs.~\cite{Segal:2023Effective,Nick:2023Effective,Brenes:2024Dynamics,Min:2024Phases,Garwola:2024}. 

The theory is built on top of the reaction-coordinate mapping, with the Hamiltonian $\hat{H}^{\rm RC}$ from Eq.~\eqref{eq:h_rc_tot}. The procedure is to apply the transformation $\hat{H}^{\rm RC-P} = \hat{U} \hat{H}^{\rm RC} \hat{U}^{\dagger}$,
\begin{align}
\label{eq:polaron_transformation}
\hat{U} = \exp\left[\sum_{\alpha}\frac{\lambda_{\alpha}}{\Omega_{\alpha}}\hat{S}^{\alpha}\left(\hat{a}_{\alpha}^{\dagger}-\hat{a}_{\alpha}\right)\right],
\end{align}
where $\hat{U}$ is known as the polaron transformation. In the present case, there is an important complication whenever the coupling operators $\hat{S}^{\alpha}$ do not commute, which was resolved in Ref. \cite{Garwola:2024}. Following the polaron transformation (\ref{eq:polaron_transformation}), one considers only the ground-state manifold of the reaction-coordinates in the polaron frame, to obtain an Effective Hamiltonian
\begin{align}
\hat{H}^{\rm eff} = \hat{Q}_0 \hat{U} \hat{H}^{\rm RC} \hat{U}^{\dagger} \hat{Q}_0,
\end{align}
where $\hat{Q}_0 = \ket{0} \bra{0}$ is the projector operator to the reaction-coordinates' ground-state manifold. The crucial aspect of this procedure is that it leads to a total system+baths Hamiltonian of the same type as Eq.~\eqref{eq:h_total}. We obtain \cite{Segal:2023Effective,Garwola:2024}
\begin{align}
\label{eq:h_total_effective}
\hat{H}^{\rm{eff}} & = \hat{Q}_0\hat{U}\hat{H}_{S}\hat{U}^{\dagger} \hat{Q}_0-\sum_{\alpha}\frac{\lambda_{\alpha}^{2}}{\Omega_{\alpha}}\left[\left(\hat{S}^{\alpha}\right)^2\right]_{\rm eff} \nonumber \\
&+\sum_{\alpha k} \left[\omega_{k,\alpha}\hat{b}_{k,\alpha}^{\dagger}\hat{b}_{k,\alpha}-\frac{2\lambda_{\alpha}f_{k,\alpha}}{\Omega_{\alpha}}\hat{S}^{\alpha}_{\rm eff}(\hat{b}_{k,\alpha}^{\dagger}+\hat{b}_{k,\alpha})\right].
\end{align}
In Eq.~\eqref{eq:h_total_effective}, we recognize the effective {\em system} Hamiltonian as
\begin{align}
\label{eq:h_sys_effective}
\hat{H}^{\rm{eff}}_S = \hat{Q}_0\hat{U}\hat{H}_{S}\hat{U}^{\dagger} \hat{Q}_0 - \sum_{\alpha}\frac{\lambda_{\alpha}^{2}}{\Omega_{\alpha}}\left[\left(\hat{S}^{\alpha}\right)^2\right]_{\rm eff}.
\end{align}
The endeavor then becomes to evaluate both $\hat{H}^{\rm eff}_S$ and $\hat{S}^{\alpha}_{\rm eff}$ explicitly, so one may employ these results to understand the underlying strong-coupling effects from the perspective of a weak-coupling theory. Note that the second term in the above equation contributes a constant in the present study.

In general, the Effective Hamiltonian theory embeds the effects of strong coupling onto both system and system-bath coupling operators.

\begin{figure}[t]
\fontsize{13}{10}\selectfont 
\centering
\includegraphics[width=\columnwidth]{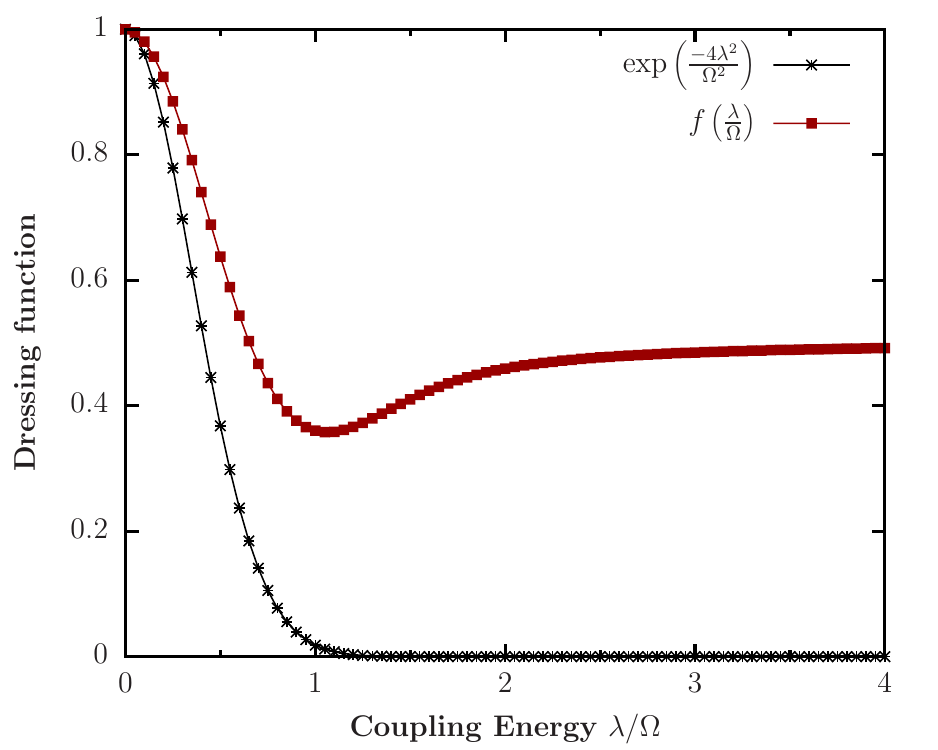}
\caption{Dressing functions with respect to the coupling energy $\lambda / \Omega$.}
\label{fig:4}
\end{figure}

We are interested in two cases: the commuting case in which $(\theta, \phi) = (\frac{\pi}{2},\frac{\pi}{2})$ and the noncommuting case for which we observe the optimal heat transfer at strong coupling $(\theta, \phi) = (\delta, \delta \pm \frac{\pi}{2})$. 

The former has been investigated in previous works~\cite{Segal:2023Effective,Nick:2023Effective,Brenes:2024Dynamics,Min:2024Phases} and leads to the expressions (with $\lambda = \lambda_{\alpha}$ and $\Omega = \Omega_{\alpha}$)
\begin{align}
\label{eq:h_rc_s_qubit}
\hat{H}_{S}^{\rm eff} &= e^{-\frac{4\lambda^{2}}{\Omega^{2}}}\Delta\hat\sigma^{z}, \nonumber \\
\hat{S}^{\alpha}_{\rm eff} &= \hat{S}^{\alpha} = \hat{\sigma}^{x};
\quad (\theta, \phi) = (\pi/2, \pi/2).
\end{align} 
From the Effective Hamiltonian theory we observe that the effect of strong coupling is to {\it suppress} the spin-splitting energy of the system Hamiltonian, while the system-bath coupling operator remains as in the weakly coupled case~\cite{Nick:2021RC}. This approach is very powerful in explaining the heat current behavior in Fig.~\ref{fig:3}: At vanishing $\lambda$, the current tends to zero as it is this coupling energy that mediates energetic transitions. At large $\lambda$, the spin-splitting energy vanishes, therefore inhibiting transitions from happening, which results in no heat current flowing through the system. There exists a maximum at a finite coupling $\lambda$ whenever the coupling is optimal with respect to the transitions that can be mediated by the system.

We next focus on the case where $\hat{S}^L$ and $\hat{S}^R$ do not commute. The calculation involved is more complicated in this case. We show in Appendix~\ref{ap:2} that (with $\lambda = \lambda_{\alpha}$ and $\Omega = \Omega_{\alpha}$)
\begin{align}
\label{eq:h_rc_s_qubit_angle}
\hat{H}_{S}^{\rm eff} &= f\left( \frac{\lambda}{\Omega} \right) \Delta\hat\sigma^{z}, \nonumber \\
\hat{S}^{\alpha}_{\rm eff} &= f\left( \frac{\lambda}{\Omega} \right) \hat{S}^{\alpha}, \quad \forall (\theta, \phi) = \left(\delta, \delta \pm \frac{\pi}{2}\right),
\end{align}
where $f(x) = 1 - x\sqrt{2}F(x\sqrt{2})$ and $F(y) = e^{-y^2}\int_0^y {\rm d}t e^{t^2}$ is the Dawson integral.

In Fig.~\ref{fig:4} we display both dressing functions, i.e., the prefactors in front of the system Hamiltonian that embed the effects of strong coupling onto the system for different choices of coupling operators. Whenever the coupling operators $\hat{S}_L = \hat{S}_R$, we obtain an exponential decay that explains the maximum and the turnover observed in the heat current (Fig.~\ref{fig:3}, black curve). 
As mentioned earlier, it is related to the vanishing spin-splitting at large $\lambda$. For the noncommuting case, for which we have chosen the points at which the heat current attains its maximum values $(\theta, \phi) = (\delta, \delta \pm \frac{\pi}{2})$, the dressing function illustrates that neither the spin-splitting energy nor the coupling operators themselves vanish in the limit of large $\lambda$. 
This would suggest that the heat current remains finite in the limit $\lambda \to \infty$, in apparent contradiction to our findings in Fig.~\ref{fig:3} (green and red curves). We investigate this further in the next section. 

Irrespective of this, the Effective Hamiltonian theory for noncommuting operators reveals an underlying symmetry that occurs whenever $(\theta, \phi) = (\delta, \delta \pm \frac{\pi}{2})$, such that both system Hamiltonian and system-bath coupling operators get dressed {\it with the same dressing function}. 

\subsection{Currents from the Effective Hamiltonian theory}
\label{sec:currents_effective}

Armed with the Effective Hamiltonian formalism, we attempt to understand strong-coupling thermodynamics from the perspective of a weak-coupling theory.

\begin{figure}[t]
\fontsize{13}{10}\selectfont 
\centering
\includegraphics[width=\columnwidth]{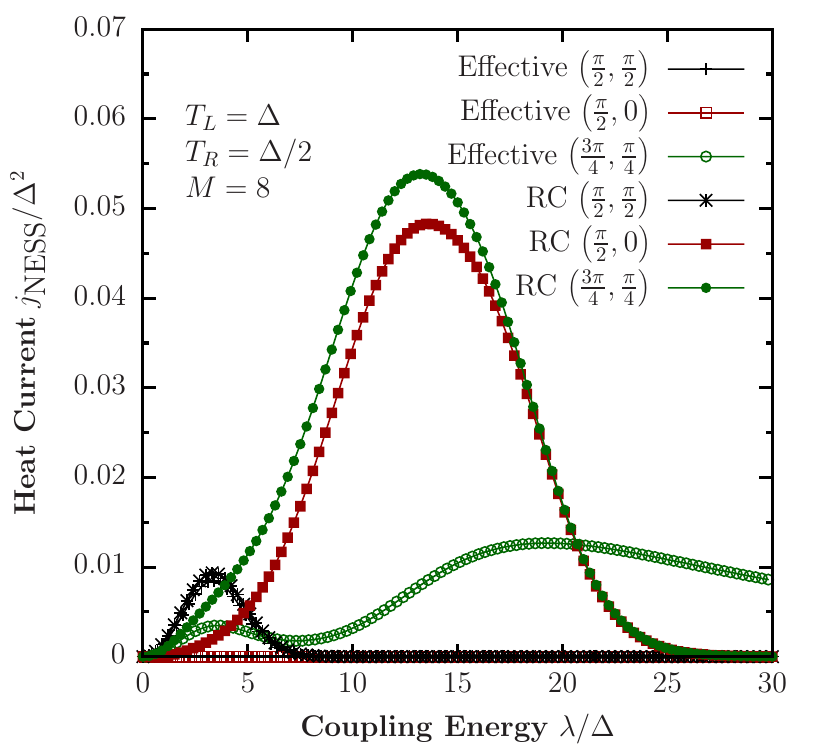}
\caption{Heat current in the NESS flowing from left ($T_L = \Delta$) towards right bath ($T_R = \Delta / 2$) as a function of system-bath coupling $\lambda_L = \lambda_R = \lambda$ for three different interaction parameters $(\theta, \phi) = (\frac{\pi}{2}, \frac{\pi}{2})$, $(\theta, \phi) = (\frac{\pi}{2}, 0)$ and $(\theta, \phi) = (\frac{3\pi}{4}, \frac{\pi}{4})$. In each case we display the reaction-coordinate calculations as in Fig.~\ref{fig:3} and those predicted by our Effective Hamiltonian model. For this calculation we used $\Omega_L = \Omega_R = \Omega = 8\Delta$, $M = 8$ and $\gamma_L = \gamma_R = 0.05 / \pi$.}
\label{fig:5}
\end{figure}

For the heat current, the procedure is the same as highlighted in Sec.~\ref{sec:ness} and Sec.~\ref{sec:strong}, with the exception that now we use the Effective Hamiltonian, Eq.~\eqref{eq:h_total_effective}, which involves both an effective system Hamiltonian and effective system-bath coupling operators. We remark that this is done in such a way that the Effective Hamiltonian introduces the effects of strong coupling while the entire system+baths configuration remains at weak coupling via our Markovian embedding.

Figure~\ref{fig:5} displays the heat current as a function of the system-bath coupling energy $\lambda = \lambda_L = \lambda_R$ for three different interaction operators with angles $(\theta, \phi) = (\frac{\pi}{2}, \frac{\pi}{2})$, $(\theta, \phi) = (\frac{\pi}{2}, 0)$ and $(\theta, \phi) = (\frac{3\pi}{4}, \frac{\pi}{4})$. For this calculation, we chose $T_L = \Delta$ and $T_R = \Delta / 2$. In Fig.~\ref{fig:5} we display the reaction-coordinate results and their corresponding Effective Hamiltonian counterparts. We start by remarking that as the mean temperature and the temperature gradient change, the interaction parameters $(\theta, \phi)$ now introduce another layer of parameters that need to be optimized to achieve the optimal heat current. From Fig.~\ref{fig:5} we observe that $(\theta, \phi) = (\frac{3\pi}{4}, \frac{\pi}{4})$ produces a higher value of the maximum current than the model $(\theta, \phi) = (\frac{\pi}{2}, 0)$, although in both cases the condition $(\theta, \phi) = (\delta, \delta \pm \frac{\pi}{2})$ is satisfied. Irrespective of this, in both cases we observe similar trends for the heat current as a function of $\lambda$, as in Fig.~\ref{fig:2} and Fig.~\ref{fig:3}.

With respect to the heat current predictions of the effective model, we see a stark contrast depending on the chosen system-bath operators $\hat{S}^{\alpha}$. For the case $(\theta, \phi) = (\frac{\pi}{2}, \frac{\pi}{2})$, the effective model correctly reproduces the heat current of the full reaction-coordinate model and therefore provides an intuitive analytical understanding of the relevant strong-coupling thermodynamics. The approximation remains valid as long as the condition $\Omega \gtrsim \lambda,T$ is satisfied~\cite{Segal:2023Effective}. The intuition behind this is that when $\Omega \lesssim \lambda,T$, additional important transport channels contribute, which are {\it not} accounted for with the effective model. This limitation arises because the model relies on a ground-state manifold truncation of the reaction coordinate in the polaron frame.
However, when $[\hat{S}^L, \hat{S}^R] \neq 0$, we observe in Fig.~\ref{fig:5} that our effective treatment does not adequately replicate the behavior of the heat current. This is an indication that the transport channels are starkly different depending on the chosen system-bath coupling operators and their commutativity. 
In the dissipative-decoherence model $(\theta, \phi) = (\frac{\pi}{2}, 0)$, the Effective Hamiltonian predicts no current flowing through the system, which we derive in Appendix \ref{ap:2}. 
This stands in complete contrast to reaction coordinate simulations. Second,
for the model $(\theta, \phi) = (\frac{3\pi}{4}, \frac{\pi}{4})$, the Effective Hamiltonian method provides two peaks, albeit the second one missing the height and position of reaction-coordinate simulations.

\begin{figure}[t]
\fontsize{13}{10}\selectfont 
\centering
\includegraphics[width=\columnwidth]{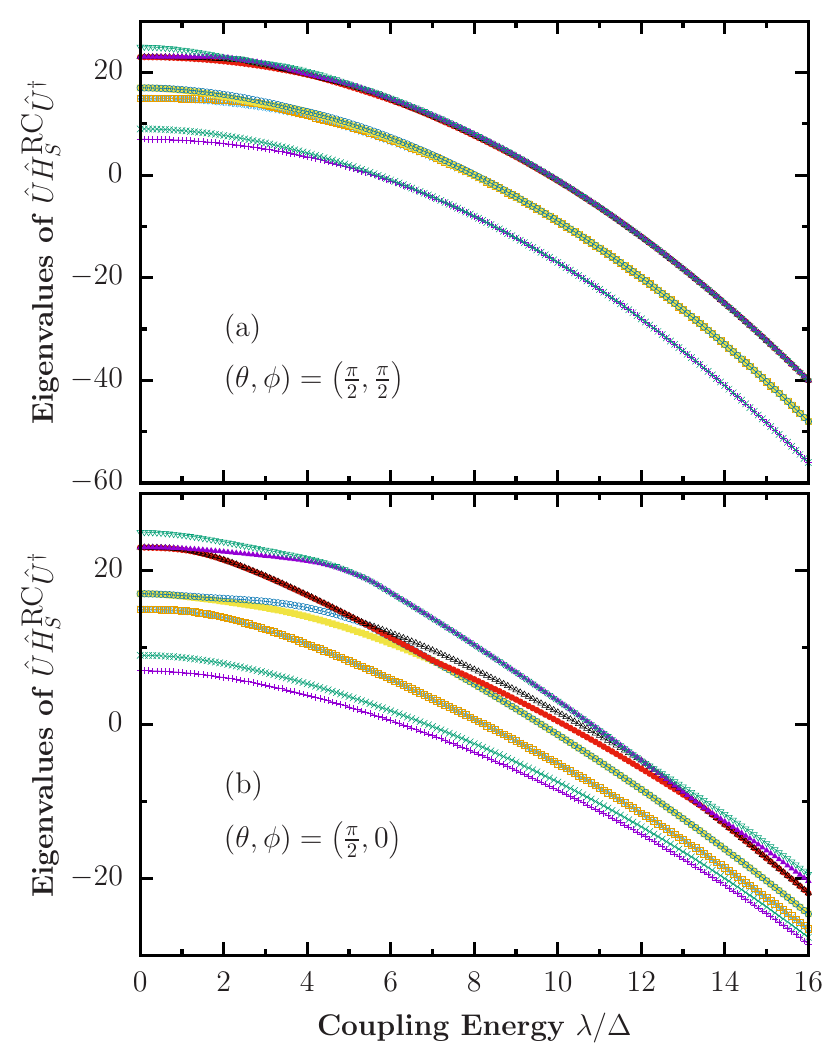}
\caption{First 10 eigenvalues of the system reaction-coordinate Hamiltonian $\hat{H}^{\rm{RC}}_S$ [Eq.~\eqref{eq:h_rc_s}] transformed by the unitary operator $\hat{U}$ [Eq.~\eqref{eq:polaron_transformation}] as a function of the system-bath coupling energy $\lambda$. Two cases are displayed. (a) The eigenvalues for commuting system-bath operators with interaction parameters $(\theta, \phi) = (\frac{\pi}{2}, \frac{\pi}{2})$ and (b) shows the noncommuting case $(\theta, \phi) = (\frac{\pi}{2}, 0)$.}
\label{fig:6}
\end{figure}
To understand the failure of the Effective Hamiltonian theory for studying heat currents under noncommuting operators,
in Fig.~\ref{fig:6} we display the first few eigenvalues of the polaron-transformed reaction coordinate Hamiltonian, $\hat{H}^{\rm{RC-P}}_S = \hat{U} \hat{H}^{\rm{RC}}_S \hat{U}^{\dagger}$ defined from Eq.~\eqref{eq:h_rc_s} and Eq.~\eqref{eq:polaron_transformation}. For this calculation, we approximate the polaron transformation by truncating the reaction coordinate modes to $M=40$, then looking at the first few eigenvalues as shown in Fig.~\ref{fig:6}. We have confirmed that the eigenvalues have converged with this value of $M$. Fig.~\ref{fig:6} reveals an important aspect when considering the case $(\theta, \phi) = (\frac{\pi}{2}, \frac{\pi}{2})$. In this model, the first two levels do not cross excited states and their gap to higher levels remains fixed as $\lambda$ increases. This justifies the truncation made in the Effective Hamiltonian treatment, as long as $\Omega \gtrsim \lambda,T$.
As a result, we obtain a good approximation for the heat current as observed in Fig.~\ref{fig:5}. On the other hand, we observe that when $[\hat{S}^L, \hat{S}^R] \neq 0$, multiple excited states become relevant once $\lambda$ increases, given that the gap between levels is reduced, even leading to level crossing. These results are an indication that an Effective Hamiltonian model, which considers only the ground-state manifold of the polaron-transformed reaction-coordinate Hamiltonian, does not provide a full picture of the transport properties for $\lambda/\Delta\gtrsim 4$.

In terms of transport mechanisms, for the family of models with ($\theta$,$\phi$) = ($\delta$,$\delta \pm \pi/2$), the current from the reaction coordinate simulations attains its maximum for $\lambda \approx 2\Omega$. We recall that $\Omega$ corresponds to the characteristic frequency of the bath, translating to the frequency of the reaction coordinate. The heat current peaks at a coupling energy value of approximately that frequency, suggesting that heat exchange is dominated by processes that exchange this amount of energy.
In some cases, one can also discern a small ``hump" at $\lambda\approx 2\Delta$ (see Fig.~\ref{fig:5}), in addition to the strong peak at $\lambda \approx 2\Omega$. Our interpretation is that there are two transport channels at play, a transport channel utilizing the spin gap, and a higher-order transport channel with a direct energy transfer between the baths, mediated by the spin system, but without a physical population transfer to it.
Importantly (see Appendix \ref{ap:2}), the Effective Hamiltonian framework is only capable in capturing transport through the central spin, and it cannot capture processes at higher frequencies, which take advantage of the reaction coordinate modes. 
The two-peak structure that the Effective Hamiltonian generates in Fig. \ref{fig:5} emerges due to the nonmonotonicity of the dressing functions of the spin splitting and coupling operators, rather than the competition of two transport channels (see Fig.~\ref{fig:4}).

With respect to quantum heat transfer in spin-boson models, we conclude that the Effective Hamiltonian method captures well the commuting case investigated here, even at strong coupling. For noncommuting operators, the dressing functions in the Effective Hamiltonian reflects the symmetry of transport with respect to $\pi/2$ shift in the angles of the coupling operators, Eq.~\eqref{eq:h_rc_s_qubit_angle}.
However, due to the truncation to the ground state manifold of the reaction coordinates, the Effective Hamiltonian technique does not capture thermal transport through higher transitions than the central spin's gap. These processes dominate transport in noncommuting models at strong coupling.
The method thus provides quantitative correct predictions only up to $\lambda \approx \Delta$, but is expected to perform better at low temperatures when transport is controlled by low-energy processes.

It is interesting to note that the Effective Hamiltonian approach showed {\it success} in capturing the {\it equilibrium state} of a spin system coupled to noncommuting baths~\cite{Garwola:2024}. It is clear that the NESS poses a new challenge to the Effective Hamiltonian method when noncommuting operators are at play, calling for future developments.

\section{Conclusions}
\label{sec:conc}

We conducted a numerical investigation of quantum heat transfer between two bosonic reservoirs mediated by a central spin degree of freedom. Our focus was on optimizing the heat current by manipulating the system-bath coupling operators. The study explored a broad range of coupling regimes, spanning from weak to ultrastrong interactions, while allowing the coupling operators to the two reservoirs to be noncommuting. Jointly, these features introduce significant technical challenges that are difficult to address using standard perturbative or nonperturbative methods.

To analyze this setup, we employed the reaction coordinate Markovian embedding method. The key finding of our study is that the optimal coupling operators for maximizing the heat current vary across different interaction regimes. In the weak coupling regime, the heat current is maximized when the central spin couples to the baths through identical operators that facilitate sequential-resonant transport, specifically $\hat{\sigma}^x$. In contrast, in the strong coupling regime, the heat current is optimized using a specific arrangement of noncommuting coupling operators. Specifically, in this regime, the system should couple to the two baths with a combination of spin components, $\hat{\sigma}^x$ and $\hat{\sigma}^z$, with their relative weight dictated by angle measures $\delta$ for one contact and $\delta\pm \pi/2$ for the other.

The Effective Hamiltonian framework was successfully used in several previous studies to capture nonperturbative system-bath coupling phenomena, and even in the ultrastrong coupling limit.
Here, we learned that one should be cautious with the Effective Hamiltonian treatment, as there are models where it critically fails to capture the relevant NESS behavior. 
Future work will be devoted to extending this treatment to allow its application on a broader range of models. 

Our optimization of the heat current was carried out under the assumption of a relatively small temperature bias across the junction. It is important to emphasize the richness of the transport model: evaluating the heat current under a high bias may yield a different set of coupling operators optimized for maximizing the current in that regime. Furthermore, we limited ourselves to the case of identical contacts by using the same spectral functions $J(\omega)$ and coupling energies $\lambda$. Introducing asymmetry at the contacts could provide additional control over both the magnitude and behavior of the quantum heat current in nanoscale junctions. Finally, it would be interesting to further examine the current noise in these nonequilibrium devices and identify setups of high current with low noise levels.

\begin{acknowledgments}
D.S. acknowledges David Gelbwaser-Klimovsky for discussions spurring research on the topic of optimizing heat transport in nanojunctions.
The work of M.B.~has been supported by an NSERC Discovery grant of D.S and Universidad de Costa Rica (Project No. Pry01-1750-2024). 
D.S.~acknowledges support from NSERC and the Canada Research Chair program. Computations were performed on the Niagara supercomputer at the SciNet HPC Consortium. SciNet is funded by: the Canada Foundation for Innovation; the Government of Ontario; the Ontario Research Fund---Research Excellence; and the University of Toronto.
This research has been funded in part by the research project: ``Quantum Software Consortium: Exploring Distributed Quantum Solutions for Canada" (QSC). QSC is financed under the National Sciences and Engineering Research Council of Canada (NSERC) Alliance Consortia Quantum Grants No. \#ALLRP587590-23.
\end{acknowledgments}

\appendix

\section{Heat transfer at stronger values of system-bath coupling energy}
\label{ap:1}

\begin{figure}[htbp]
\fontsize{13}{10}\selectfont 
\centering
\includegraphics[width=\columnwidth]{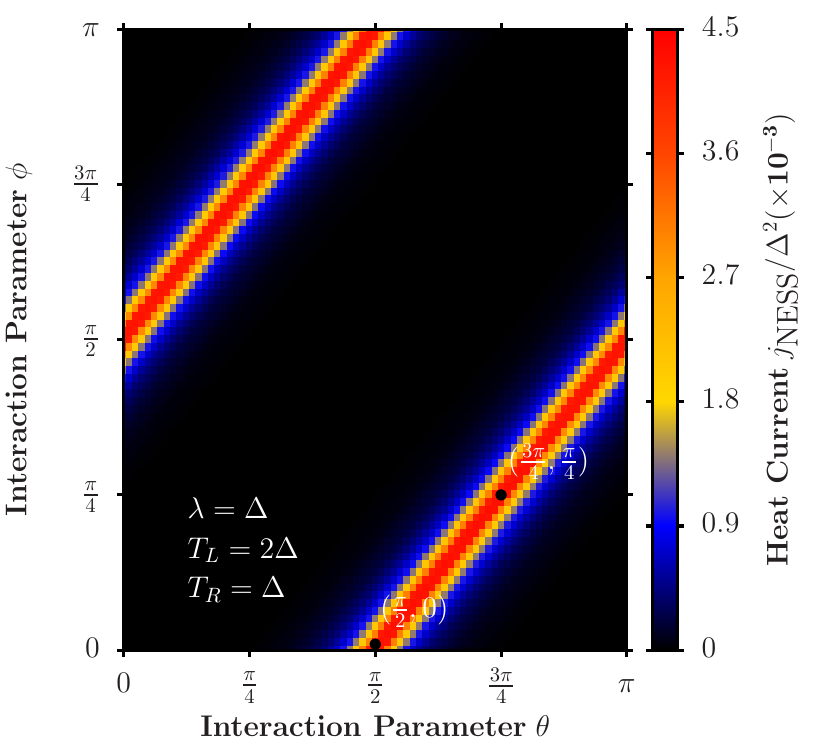}
\caption{Heat current in the NESS flowing from left ($T_L = 2\Delta$) towards right bath ($T_R = \Delta$) at strong system-bath coupling $\lambda_L = \lambda_R = \lambda = 20\Delta$ as a function of the interaction parameters $(\theta, \phi)$. For this calculation we used $\Omega_L = \Omega_R = \Omega = 8\Delta$, $M = 6$ and $\gamma_L = \gamma_R = 0.05 / \pi$. }
\label{fig:ap1}
\end{figure}

We present here an additional example of the dependence of the heat current on the two angle parameters $\theta$ and $\phi$, which control the coupling operators of the system to the baths. Extending Fig.~\ref{fig:1} and Fig.~\ref{fig:2}, 
Fig.~\ref{fig:ap1} examines the case where the coupling energy is as large as $\lambda = 20\Delta$. The results reveal that the heat current is maximized when $(\theta, \phi) = (\delta, \delta \pm \pi/2)$, but it rapidly decreases as the parameters deviate from this configuration. This behavior indicates that in the ultrastrong coupling limit, the system becomes insulating outside this symmetry.

\section{Convergence of the reaction-coordinate calculations with respect to level truncation}
\label{ap:3}

\begin{figure}[htbp]
\fontsize{13}{10}\selectfont 
\centering
\includegraphics[width=0.4\textwidth]{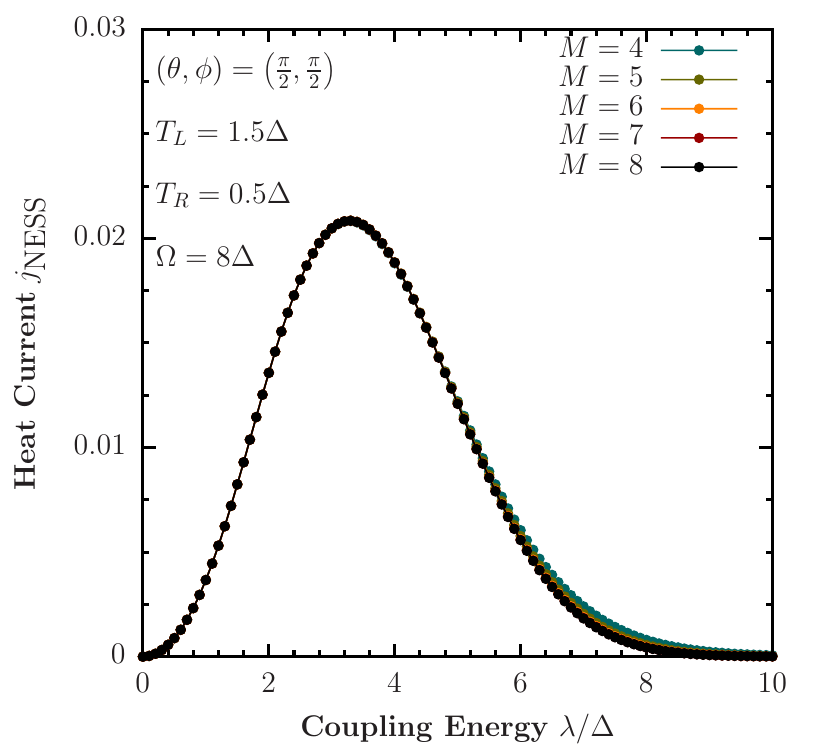}\\
\includegraphics[width=0.4\textwidth]{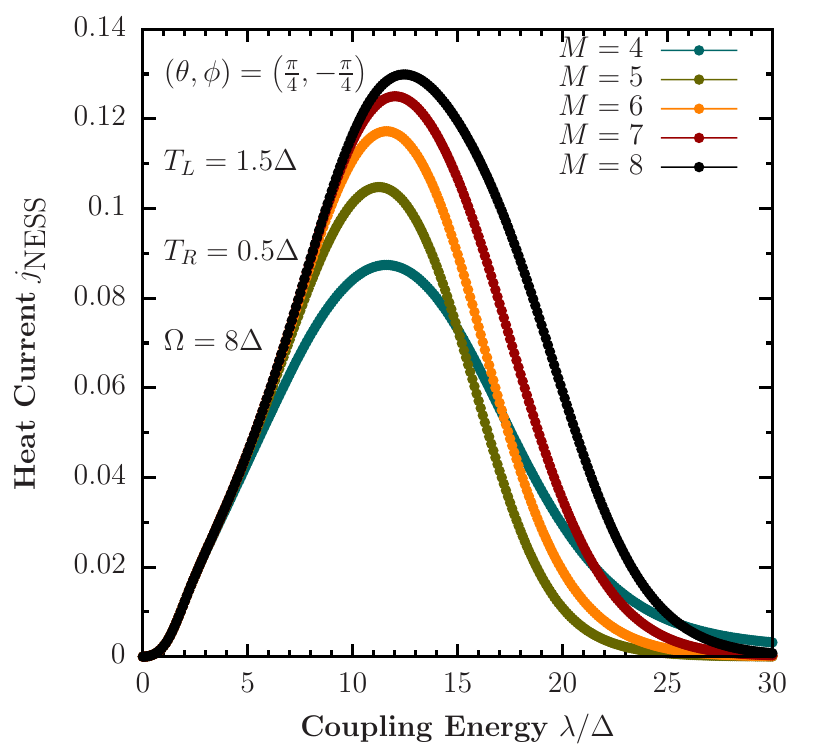}
\caption{Comparison of the NESS heat currents with respect to reaction-coordinate level truncation $M$. The heat current is flowing from the left ($T_L = 1.5\Delta$) towards the right bath ($T_R = 0.5\Delta$), and it is presented here as a function of the coupling energy $\lambda = \lambda_L = \lambda_R$ for two different interaction parameters $(\theta, \phi) = \left(\frac{\pi}{2}, \frac{\pi}{2}\right)$ (top) and $(\theta, \phi) = \left(\frac{\pi}{4}, -\frac{\pi}{4}\right)$ (bottom). For this calculation we used $\Omega_L = \Omega_R = \Omega = 8\Delta$ and $\gamma_L = \gamma_R = 0.05/ \pi$.}
\label{fig:ap2}
\end{figure}

In this Appendix, we discuss the convergence of reaction-coordinate simulations with respect to the truncation of the RC manifold.
In our model we extract two RCs, one from each bath, where for simplicity we assume that the baths are characterized by identical spectral functions with the same parameters, $\Omega$, $\gamma$ and $\lambda$, as the characteristic frequency of the baths, width of the spectral function, and coupling energy to the system, respectively. 
Given that extracted RCs are identical, we assume that similar truncations should be done on both baths, i.e., we maintain $M$ levels for each RC.

In Fig. \ref{fig:ap2}, we present simulation results. In the top panel, both baths are coupled to the system through the $\hat \sigma^x$ operator. In this model, which was previously examined in Ref.~\cite{Nick:2021RC} using the reaction coordinate method, convergence with respect to $M$ is easily achieved, with some small deviations showing up as one increases $\lambda$ to $\lambda/\Delta \geq 6$. However, in that regime, the current is strongly suppressed, and thus these deviations have a small impact on the current. The exceptional success and convergence of RC simulations can be rationalized by the analysis of level spacings, which, as illustrated in Fig. \ref{fig:6}(a), shows the robustness of the energy spectrum even at strong coupling~\cite{Nick:2021RC}.

Moving on to the bottom panel of Fig.~\ref{fig:ap2}, here we analyze the case with noncommuting system operators coupled to the two baths, with an angle difference of $\pi/2$.  Focusing on convergence trends with respect to $M$, the following is clear. (i) At weak to intermediate coupling, up to $\lambda/\Delta\approx 5$, we reach excellent convergence with $M=4$ capturing results.
(ii) Close to the maximum, at  $\lambda/\Delta\approx 10$, we note that the current is monotonically converging as we increase $M$. (iii) In the ultrastrong coupling regime, $\lambda/\Delta>20$, while all curves show similar trends (suppression of the heat current with increasing coupling energy), it is necessary to maintain more levels in each RC to converge. 

The challenge in achieving convergence under general coupling operators can once again be understood by examining the eigenenergies of the RC Hamiltonian. As shown in Fig.~\ref{fig:6}(b), increasing $\lambda$ causes energy levels to approach and cross each other. This behavior requires the inclusion of a large number of RC levels to accurately cover the relevant energetic window, which is determined by the ratio $\Omega/\Delta$ and $\Omega/T_{L,R}$.

\section{Details on the Effective Hamiltonian theory}
\label{ap:2}

In this Appendix, we present derivations of the effective system operators in different cases and derive analytical formulas for the heat current for the effective models discussed in the main text. 
While the method failed to reproduce the strong coupling trends of heat currents for general noncommuting operators, we detail it here for several reasons. First, it provides qualitative (and even quantitative) correct results beyond the strict weak coupling limit, see Fig. \ref{fig:5}.
This is because the method correctly captures transport behavior thorough the central spin---though it is missing additional transport pathways.
Second, the analytical expressions of the Effective Hamiltonian method provide us with a good understanding of the physics controlling transport at low frequencies.
Third, the effective method is expected to become more accurate at low temperatures, when transport through high energy levels is suppressed. 

The Effective Hamiltonian theory in the case of noncommuting system operators was developed in details in Ref. \cite{Garwola:2024}. We present here the technical steps relevant for our model.

\begin{widetext}
We need to evaluate the following two integrals over momentum. Each integral is two dimensional, as it covers momentum related to each reaction coordinate, 
\begin{equation}
\begin{aligned}(\hat{\sigma}^{z})^{\text{eff}} & =\int\hat{U}_{P}(\tilde{\boldsymbol{p}})\hat{\sigma}^{z}\hat{U}_{P}^{\dagger}(\tilde{\boldsymbol{p}})\prod_{\alpha=L,R}\frac{e^{-\tilde{p}_{\alpha}^{2}}d\tilde{p}_{\alpha}}{\sqrt{\pi}},\\
(\hat{\sigma}^{x})^{\text{eff}} & =\int\hat{U}_{P}(\tilde{\boldsymbol{p}})\hat{\sigma}^{x}\hat{U}_{P}^{\dagger}(\tilde{\boldsymbol{p}})\prod_{\alpha=L,R}\frac{e^{-\tilde{p}_{\alpha}^{2}}d\tilde{p}_{\alpha}}{\sqrt{\pi}}.
\label{eq:Asigeff}
\end{aligned}
\end{equation}
For coupling operators defined by the following two angle parameters, 
$\hat{S}^L = \cos(\theta) \hat{\sigma}^z + \sin(\theta) \hat{\sigma}^x$ and 
$\hat{S}^R = \cos(\phi) \hat{\sigma}^z + \sin(\phi) \hat{\sigma}^x$, 
the unitary polaron transformation is given by
\begin{equation}
\label{eq:AUpol}
\begin{aligned}\hat{U}_{P}(\boldsymbol{\tilde{p}}) & =\exp\left(-i\sqrt{2}\frac{\lambda}{\Omega}\sum_{\alpha}\tilde{p}_{\alpha}\hat{S}^{\alpha}\right)\\
 & =\hat{\sigma}^{0}\cosh\left(\sqrt{2}\frac{\lambda}{\Omega}\sqrt{-2\tilde{p}_{L}\tilde{p}_{R}\cos(\theta-\phi)-\tilde{p}_{L}^{2}-\tilde{p}_{R}^{2}}\right)\\
 & -\hat{\sigma}^{x}\frac{i\left(\sin(\theta)\tilde{p}_{L}+\tilde{p}_{R}\sin(\phi)\right)\sinh\left(\sqrt{2}\frac{\lambda}{\Omega}\sqrt{-2\tilde{p}_{L}\tilde{p}_{R}\cos(\theta-\phi)-\tilde{p}_{L}^{2}-\tilde{p}_{R}^{2}}\right)}{\sqrt{-2\tilde{p}_{L}\tilde{p}_{R}\cos(\theta-\phi)-\tilde{p}_{L}^{2}-\tilde{p}_{R}^{2}}}\\
 & -\hat{\sigma}^{z}\frac{i\left(\cos(\theta)\tilde{p}_{L}+\tilde{p}_{R}\cos(\phi)\right)\sinh\left(\sqrt{2}\frac{\lambda}{\Omega}\sqrt{-2\tilde{p}_{L}\tilde{p}_{R}\cos(\theta-\phi)-\tilde{p}_{L}^{2}-\tilde{p}_{R}^{2}}\right)}{\sqrt{-2\tilde{p}_{L}\tilde{p}_{R}\cos(\theta-\phi)-\tilde{p}_{L}^{2}-\tilde{p}_{R}^{2}}}.
\end{aligned}
\end{equation}
This expression was derived using $\lambda=\lambda_{L,R}$ and $\Omega=\Omega_{L,R}$.

\subsection{Dressing of operators in the Effective Hamiltonian framework for $(\theta,\phi)=(\delta,\delta\pm\frac{\pi}{2})$}

We now focus on the special case of $(\theta,\phi)=(\delta,\delta\pm\frac{\pi}{2})$. In this case, the polaron operator simplifies to 
\begin{equation}
\begin{aligned}\hat{U}_{P}(\boldsymbol{\tilde{p}}) & =\hat{\sigma}^{0}\cos\left(\sqrt{2}\frac{\lambda}{\Omega}\sqrt{\tilde{p}_{L}^{2}+\tilde{p}_{R}^{2}}\right)\\
 & -\hat{\sigma}^{x}\frac{i\sin\left(\sqrt{2}\frac{\lambda}{\Omega}\sqrt{\tilde{p}_{L}^{2}+\tilde{p}_{R}^{2}}\right)\left(\sin(\delta)\tilde{p}_{L}\pm\cos(\delta)\tilde{p}_{R}\right)}{\sqrt{\tilde{p}_{L}^{2}+\tilde{p}_{R}^{2}}}\\
 & -\hat{\sigma}^{z}\frac{i\sin\left(\sqrt{2}\frac{\lambda}{\Omega}\sqrt{\tilde{p}_{L}^{2}+\tilde{p}_{R}^{2}}\right)\left(\cos(\delta)\tilde{p}_{L}\mp\sin(\delta)\tilde{p}_{R}\right)}{\sqrt{\tilde{p}_{L}^{2}+\tilde{p}_{R}^{2}}}.
\end{aligned}
\end{equation}
We simplify the integrals in Eq. (\ref{eq:Asigeff}) by using the substitution $\tilde{p}_{L}=r\cos(\alpha)$,
$\tilde{p}_{R}=r\sin(\alpha)$. We arrive at
\begin{equation}
\begin{aligned}(\hat{\sigma}^{z})^{\text{eff}} & =\int_{0}^{\infty}re^{-r^{2}}dr\int_{0}^{2\pi}\frac{d\alpha}{\pi}{\Big\{}
\hat{\sigma}^{z}\Big[\cos^{2}(\sqrt{2}\frac{\lambda}{\Omega}r)
  +\cos\left(2(\alpha\pm\delta)\right)\sin^{2}(\sqrt{2}\frac{\lambda}{\Omega}r)\Big]\\
 & +\hat{\sigma}^{x}\Big[\pm\sin\left(2(\alpha\pm\delta)\right)\sin^{2}\left(\sqrt{2}\frac{\lambda}{\Omega}r\right)\Big]\\
 & +\hat{\sigma}^{y}\Big[\mp2\sin(\alpha\pm\delta)\sin\left(\sqrt{2}\frac{\lambda}{\Omega}r\right)\Big]{ \Big\}},\\
(\hat{\sigma}^{x})^{\text{eff}} & =\int_{0}^{\infty}re^{-r^{2}}dr\int_{0}^{2\pi}\frac{d\alpha}{\pi}
{ \Big\{}
\hat{\sigma}^{z}\left[\pm\sin\left(2(\alpha\pm\delta)\right)\sin^{2}\left(\sqrt{2}\frac{\lambda}{\Omega}r\right)\right]\\
 & +\hat{\sigma}^{x}\left[\cos^{2}\left(\sqrt{2}\frac{\lambda}{\Omega}r\right)-\cos\left(2(\alpha\pm\delta)\right)\sin^{2}(\sqrt{2}\frac{\lambda}{\Omega}r)\right]\\
 & +\hat{\sigma}^{y}\left[\cos(\alpha\pm\delta)\sin\left(2\sqrt{2}\frac{\lambda}{\Omega}r\right)\right]
 { \Big\}}.
\end{aligned}
\end{equation}
Integration over the angle $\alpha$ greatly simplifies the above expressions, reaching
\begin{equation}
\begin{aligned}(\hat{\sigma}^{z})^{\text{eff}} & =\hat{\sigma}^{z}\int_{0}^{\infty}2e^{-r^{2}}r\cos^{2}\left(\sqrt{2}\lambda r\right)dr,\,\,\,\,\,\,\,\,\,
(\hat{\sigma}^{x})^{\text{eff}} & =\hat{\sigma}^{x}\int_{0}^{\infty}2e^{-r^{2}}r\cos^{2}\left(\sqrt{2}\lambda r\right)dr.
\end{aligned}
\end{equation}
\end{widetext}
We can express the above integrals as Dawson functions,
as presented in the main text, below Eq. (\ref{eq:h_rc_s_qubit_angle}).

These dressing of operators impact the Hamiltonian in two substantial ways.
First, the dressing of $\hat\sigma^z$ translates to dressing of the spin energy splitting $\Delta$, as demonstrated in Eq. (\ref{eq:h_rc_s_qubit_angle}) and in Fig. \ref{fig:4}. The dressing is nonmonotonic, but generally, its impact is to reduce $\Delta$ to ultimately $\Delta /2$ at the ultrastrong coupling limit. 
The second impact of the dressing is to modify the system-bath coupling energies.
These modification is symmetric-identical when the spectral functions are assumed the same; see Eq. (\ref{eq:h_rc_s_qubit_angle}).
Both these aspects should impact the heat current, as we show in the next subsection. Particularly, a suppression of the splitting should reduce the sequential heat current with each transition carrying less energy. 

\subsection{Heat current in the effective Hamiltonian framework}

We proceed to find the analytical expressions for the steady state heat current using the Redfield equation (\ref{eq:redfield_dyn}).
Once we find the steady state population from the Redfield equation, we compute the heat current, flowing between the system and the left bath,
\begin{equation}
j_{\text{NESS}}=-2\tilde{\Delta}\Big(k_{L}^{\downarrow}\rho_{22}^{\text{SS}}-k_{L}^{\uparrow}\rho_{11}^{\text{SS}}\Big).\label{eq:QNESS_general}
\end{equation}
According to our sign convention, the heat current is positive when flowing into the system.
Here, $\tilde{\Delta}$ is the suppressed system energy splitting,
$k_{L}^{\downarrow}$ and $k_{L}^{\uparrow}$ are the rates of transition
from the excited to the ground state of the system and vice versa. These rate stand as a general notation.

Below, we derive the explicit expressions for these rates for each model in the language of the Redfield tensor.

It is sufficient to find the heat current between the system and the left bath, since in the NESS it is equal to minus the heat current between the system and the right bath. We focus here on deriving analytical expressions for the heat
current for models examined in Sec.~\ref{sec:effective}. 

\begin{widetext}

\subsubsection{The commuting model, $(\phi,\theta)=(\frac{\pi}{2},\frac{\pi}{2})$}

We focus on the case $(\theta,\phi)=(\frac{\pi}{2},\frac{\pi}{2})$.
The Redfield equations for the diagonal elements of the reduced density matrix are given by 
\begin{equation}
\begin{aligned}\frac{d}{dt}\rho_{11}(t) & =-\sum_{\alpha=L,R}\Big([R_{12,21}^{\alpha,\text{eff}}(\omega_{12})+R_{12,21}^{\alpha,*,\text{eff}}(\omega_{12})]\rho_{11}(t)-\big[R_{21,12}^{\alpha,\text{eff}}(\omega_{21})+R_{21,12}^{\alpha,*,\text{eff}}(\omega_{21})\big]\rho_{22}(t)\Big)\\
= & -2\sum_{\alpha=L,R}\text{Re}\Big([R_{12,21}^{\alpha,\text{eff}}(\omega_{12})+R_{21,12}^{\alpha,\text{eff}}(\omega_{21})]\rho_{11}(t)-R_{21,12}^{\alpha,\text{eff}}(\omega_{21})\Big),
\end{aligned}
\end{equation}
and
\begin{equation}
\begin{aligned}\frac{d}{dt}\rho_{22}(t) & =-\sum_{\alpha=L,R}\Big([R_{21,12}^{\alpha,\text{eff}}(\omega_{21})+R_{21,12}^{\alpha,*,\text{eff}}(\omega_{21})]\rho_{22}(t)-\big[R_{12,21}^{\alpha,\text{eff}}(\omega_{12})+R_{12,21}^{\alpha,*,\text{eff}}(\omega_{12})\big]\rho_{11}(t)\Big)\\
 & =-2\sum_{\alpha=L,R}\text{Re}\Big([R_{21,12}^{\alpha,\text{eff}}(\omega_{21})+R_{12,21}^{\alpha,\text{eff}}(\omega_{12})]\rho_{22}(t)-R_{12,21}^{\alpha,\text{eff}}(\omega_{12})\Big).
\end{aligned}
\end{equation}
We readily find the steady state diagonal elements of the reduced density matrix,
\begin{equation}
\begin{aligned}\rho_{11}^{SS} & =\frac{\sum_{\alpha=L,R}\text{Re}\Big(R_{21,12}^{\alpha,\text{eff}}(\omega_{21})\Big)}{\sum_{\alpha=L,R}\text{Re}\Big([R_{12,21}^{\alpha,\text{eff}}(\omega_{12})+R_{21,12}^{\alpha,\text{eff}}(\omega_{21})]\Big)},\end{aligned}
\end{equation}
and
\begin{equation}
\begin{aligned}\rho_{22}^{SS} & =\frac{\sum_{\alpha=L,R}\text{Re}\Big(R_{12,21}^{\alpha,\text{eff}}(\omega_{12})\Big)}{\sum_{\alpha=L,R}\text{Re}\Big([R_{21,12}^{\alpha,\text{eff}}(\omega_{21})+R_{12,21}^{\alpha,\text{eff}}(\omega_{12})]\Big)}.\end{aligned}
\end{equation}
The steady state heat current is equal to 
\begin{equation}
\label{eq:jM1}
\begin{aligned}j_{\text{NESS}} & =-4\tilde{\Delta}\text{Re}\Big(R_{12,21}^{L,\text{eff}}(\omega_{21})\rho_{22}^{\text{SS}}-R_{21,12}^{L,\text{eff}}(\omega_{12})\rho_{11}^{\text{SS}}\Big).\end{aligned}
\end{equation}
We assume that $\lambda_{L}=\lambda_{R}=\lambda$.
In the present model, the dressing leads to 
$\tilde{\Delta}=\Delta\exp(-4\frac{\lambda^{2}}{\Omega^{2}})$. Furthermore, the elements in the dissipator are, for both $\alpha=L,R$,
\begin{equation}
\begin{aligned}R_{12,21}^{\alpha,\text{eff}}(\omega_{12}) & =R_{21,12}^{\alpha,\text{eff}}(\omega_{12}),\\
 & {\rm Re}\left(R_{12,21}^{\alpha,\text{eff}}(\omega_{12})\right)=\Gamma_{\text{eff}}^{\alpha}(\omega_{12})=\pi J_{\text{eff}}^{\alpha}(2\tilde{\Delta})n^{\alpha}(2\tilde{\Delta}),\\
R_{12,21}^{\alpha,\text{eff}}(\omega_{21}) & =R_{21,12}^{\alpha,\text{eff}}(\omega_{21}),\\
 & 
 {\rm Re}\left(R_{12,21}^{\alpha,\text{eff}}(\omega_{21})\right)=\Gamma_{\text{eff}}^{\alpha}(\omega_{21})=\pi J_{\text{eff}}^{\alpha}(2\tilde{\Delta})\big(n^{\alpha}(2\tilde{\Delta})+1\big),
\end{aligned}
\end{equation}
where we define
\bea
J_{\text{eff}}^{\alpha}(\omega)=\frac{4\lambda^2}{\Omega^2}J_{\text {RC}}^\alpha(\omega).
\label{eq:Jeff}
\eea 
Altogether, plugging the steady state populations and the rate constants in Eq. (\ref{eq:jM1}), we obtain the heat current

\begin{equation}
\label{eq:C12}
\begin{aligned}j_{\text{NESS}} 
 & =-4\pi\tilde{\Delta}\left[\frac{J_{\text{eff}}^{L}(2\tilde{\Delta})\big(n^{L}(2\tilde{\Delta})+1\big)\sum_{\alpha=L,R}J_{\text{eff}}^{\alpha}(2\tilde{\Delta})n^{\alpha}(2\tilde{\Delta})-J_{\text{eff}}^{L}(2\tilde{\Delta})n^{L}(2\tilde{\Delta})\sum_{\alpha=L,R}J_{\text{eff}}^{\alpha}(2\tilde{\Delta})\big(n^{\alpha}(2\tilde{\Delta})+1\big)}{\sum_{\alpha=L,R}J_{\text{eff}}^{\alpha}(2\tilde{\Delta})\big(2n^{\alpha}(2\tilde{\Delta})+1\big)}\right]\\
 & =-4\pi\tilde{\Delta}\left[\frac{J_{\text{eff}}^{L}(2\tilde{\Delta})J_{\text{eff}}^{R}(2\tilde{\Delta})\Big(n^{R}(2\tilde{\Delta})\big(n^{L}(2\tilde{\Delta})+1\big)-n^{L}(2\tilde{\Delta})\big(n^{R}(2\tilde{\Delta})+1\big)\Big)}{\sum_{\alpha=L,R}J_{\text{eff}}^{\alpha}(2\tilde{\Delta})\big(2n^{\alpha}(2\tilde{\Delta})+1\big)}\right]\\
 & =\frac{4\pi\tilde{\Delta}J_{\text{eff}}^{L}(2\tilde{\Delta})J_{\text{eff}}^{R}(2\tilde{\Delta})\left[n^{L}(2\tilde{\Delta})-n^{R}(2\tilde{\Delta})\right]}{\sum_{\alpha=L,R}J_{\text{eff}}^{\alpha}(2\tilde{\Delta})\big[2n^{\alpha}(2\tilde{\Delta})+1\big]}.
\end{aligned}
\end{equation}
This result was originally derived in Ref. \cite{Segal:2005,Segal:2006} at weak coupling, and later presented for the effective framework in Ref. \cite{Nick:2021RC}.

\subsubsection{The dissipative-decoherence model, $(\theta,\phi)=(\frac{\pi}{2},0)$}

In the case where $(\theta,\phi)=(\frac{\pi}{2},0)$, the nonzero elements of the Redfield tensor are different for the left and right baths. The equations of motion for the populations are
\begin{equation}
\begin{aligned}\frac{d}{dt}\rho_{11}(t) & =-2\text{Re}\Big(R_{12,21}^{L,\text{eff}}(\omega_{12})\rho_{11}(t)-R_{21,12}^{L,\text{eff}}(\omega_{21})\rho_{22}(t)\Big)\\
 & \;-2\text{Re}\Big(R_{11,11}^{R,\text{eff}}(\omega_{11})\rho_{11}(t)-R_{11,11}^{R,\text{eff}}(\omega_{11})\rho_{11}(t)\Big)\\
 & =-2\text{Re}\Big([R_{12,21}^{L,\text{eff}}(\omega_{12})+R_{21,12}^{L,\text{eff}}(\omega_{21})]\rho_{11}(t)-R_{21,12}^{L,\text{eff}}(\omega_{21})\Big),
\end{aligned}
\end{equation}
and
\begin{equation}
\begin{aligned}\frac{d}{dt}\rho_{22}(t) & =-2\text{Re}\Big(R_{21,12}^{L,\text{eff}}(\omega_{21})\rho_{22}(t)-R_{12,21}^{L,\text{eff}}(\omega_{12})\rho_{11}(t)\Big)\\
 & \;-2\text{Re}\Big(R_{22,22}^{R,\text{eff}}(\omega_{22})\rho_{22}(t)-R_{22,22}^{R,\text{eff}}(\omega_{22})\rho_{22}(t)\Big)\\
 & =-2\text{Re}\Big([R_{21,12}^{L,\text{eff}}(\omega_{21})+R_{12,21}^{L,\text{eff}}(\omega_{12})]\rho_{22}(t)-R_{12,21}^{L,\text{eff}}(\omega_{12})\Big).
\end{aligned}
\end{equation}
Again, the steady state values of the diagonal elements of $\rho$ are
\begin{equation}
\begin{aligned}\rho_{11}^{SS} & =\frac{\text{Re}\left(R_{21,12}^{L,\text{eff}}(\omega_{21})\right)}{\text{Re}\left(R_{12,21}^{L,\text{eff}}(\omega_{12})\right)+\text{Re}\left(R_{21,12}^{L,\text{eff}}(\omega_{21})\right)}, \,\,\,\,\,
\rho_{22}^{SS} & =\frac{\text{Re}\left(R_{12,21}^{L,\text{eff}}(\omega_{12})\right)}{\text{Re}\left(R_{21,12}^{L,\text{eff}}(\omega_{21})\right)+\text{Re}\left(R_{12,21}^{L,\text{eff}}(\omega_{12})\right)}.
\end{aligned}
\end{equation}
That is, since only the left bath allows population dynamics, it dictates the steady state population, while the right bath does not impact the population. 

The elements of the Redfield tensor are 
\begin{equation}
\begin{aligned}
\text{Re}\left(R_{12,21}^{L,\text{eff}}(\omega_{12}) \right)& =\text{Re}\left(R_{21,12}^{L,\text{eff}}(\omega_{12})\right)\\
 & =f^{2}\left(\frac{\lambda}{\Omega}\right)\Gamma_{\text{eff}}^{L}(\omega_{12})=\pi f^{2}\left(\frac{\lambda}{\Omega}\right)J_{\text{eff}}^{L}(2\tilde{\Delta})n^{L}(2\tilde{\Delta}),\\
\text{Re}\left(R_{12,21}^{L,\text{eff}}(\omega_{21}) \right)& =\text{Re}\left(R_{21,12}^{L,\text{eff}}(\omega_{21})\right)\\
 & =f^{2}\left(\frac{\lambda}{\Omega}\right)\Gamma_{\text{eff}}^{L}(\omega_{21})=\pi f^{2}\left(\frac{\lambda}{\Omega}\right)J_{\text{eff}}^{L}(2\tilde{\Delta})\big(n^{L}(2\tilde{\Delta})+1\big).
\end{aligned}
\end{equation}
Using $\lambda_{L}=\lambda_{R}=\lambda$, $\tilde{\Delta}=f(\frac{\lambda}{\Omega})\Delta$.
The steady state heat current is equal to 
\begin{equation}
\begin{aligned}j_{\text{NESS}} & =-4\tilde{\Delta}\text{Re}\Big(R_{12,21}^{L,\text{eff}}(\omega_{21})\rho_{22}^{\text{SS}}-R_{21,12}^{L,\text{eff}}(\omega_{12})\rho_{11}^{\text{SS}}\Big)\\
 & =0.
\end{aligned}
\end{equation}
\end{widetext}
The Effective method does not properly capture the heat current in this situation, since it is a higher order process on the reaction coordinate manifold. 
Explicitly, our spin system is coupled through $\hat\sigma^x$ at one side, which allows heat exchange at weak coupling, as resonant excitations, and through $\hat\sigma^z$ in the other side. The latter operator does not allow energy exchange with the attached bath. 
Since the Effective Hamiltonian approach does not generate additional coupling terms, it ends with zero current in this case, similarly to the original weak coupling limit. 

\begin{widetext}
\subsubsection{Models with $(\theta,\phi)=(\delta,\delta\pm\pi/2)$}

Let us repeat the same derivation for more general
case, where two angle parameters are shifted by $\frac{\pi}{2}$, 
\begin{equation}
\hat{S}^{L}=\cos(\delta)\hat{\sigma}^{z}+\sin(\delta)\hat{\sigma}^{x},\;\;\;\hat{S}^{R}=\mp\sin(\delta)\hat{\sigma}^{z}\pm\cos(\delta)\hat{\sigma}^{x},
\end{equation}
where $\delta\in(0,\pi)$.

The Redfield equations for the diagonal elements of the systems reduced density matrix are
\begin{equation}
\begin{aligned}\frac{{\rm d}\rho_{11}}{{\rm d}t} 
 & =-\sum_{\alpha}\Big[[R_{12,21}^{\alpha,\text{eff}}(\omega_{12})]^{*}\rho_{11}+R_{12,21}^{\alpha,\text{eff}}(\omega_{12})\rho_{11}-R_{21,12}^{\alpha,\text{eff}}(\omega_{21})\rho_{22}-[R_{21,12}^{\alpha,\text{eff}}(\omega_{21})]^{*}\rho_{22}\\
 & +\left([R_{1222}^{\alpha,\text{eff}}(\omega_{22})]^{*}-R_{2111}^{\alpha,\text{eff}}(\omega_{11})\right)\rho_{12}+\left(R_{1222}^{\alpha,\text{eff}}(\omega_{22})-[R_{2111}^{\alpha,\text{eff}}(\omega_{11})]^{*}\right)\rho_{21}\Big]\\
 & =-\sum_{\alpha}2\text{Re}\Big([R_{12,21}^{\alpha,\text{eff}}(\omega_{12})+R_{21,12}^{\alpha,\text{eff}}(\omega_{21})]\rho_{11}-R_{21,12}^{\alpha,\text{eff}}(\omega_{21})\Big),
\end{aligned}
\end{equation}
where in the last equality we enacted the secular approximation. Under the same approximation,
\begin{equation}
\begin{aligned}\frac{{\rm d}\rho_{22}}{{\rm d}t} & =-\sum_{\alpha}[R_{21,12}^{\alpha,\text{eff}}(\omega_{21})]^{*}\rho_{22}+R_{21,12}^{\alpha,\text{eff}}(\omega_{21})\rho_{22}-R_{12,21}^{\alpha,\text{eff}}(\omega_{12})\rho_{11}-[R_{12,21}^{\alpha,\text{eff}}(\omega_{12})]^{*}\rho_{11}\\
 & =-\sum_{\alpha}2\text{Re}\Big([R_{21,12}^{\alpha,\text{eff}}(\omega_{21})+R_{12,21}^{\alpha,\text{eff}}(\omega_{12})]\rho_{22}-R_{12,21}^{\alpha,\text{eff}}(\omega_{12})\Big).
\end{aligned}
\end{equation}
The corresponding steady state populations are 
\begin{equation}
\begin{aligned}\rho_{22}^{SS} & =\frac{\sum_{\alpha}\text{Re}R_{12,21}^{\alpha,\text{eff}}(\omega_{12})}{\text{Re}\sum_{\alpha}\left(R_{21,12}^{\alpha,\text{eff}}(\omega_{21})+R_{12,21}^{\alpha,\text{eff}}(\omega_{12})\right)},\,\,\,\,
\rho_{11}^{SS} & =\frac{\sum_{\alpha}\text{Re}R_{21,12}^{\alpha,\text{eff}}(\omega_{21})}{\text{Re}\sum_{\alpha}\left(R_{12,21}^{\alpha,\text{eff}}(\omega_{12})+R_{21,12}^{\alpha,\text{eff}}(\omega_{21})\right)}.
\end{aligned}
\end{equation}
The relevant elements of the Redfield tensor are 
\begin{equation}
\begin{aligned}
\text{Re}\left(
R_{12,21}^{L,\text{eff}}(\omega_{12})\right) & =\text{Re}\left(R_{21,12}^{L,\text{eff}}(\omega_{12})\right)\\
 & =\sin^{2}(\delta)f^{2}(\epsilon)\Gamma_{\text{eff}}^{L}(\omega_{12})=\pi\sin^{2}(\delta)f^{2}(\epsilon)J_{\text{eff}}^{L}(2\tilde{\Delta})n^{L}(2\tilde{\Delta}),\\
\text{Re}\left(R_{12,21}^{L,\text{eff}}(\omega_{21})\right) & =\text{Re}\left(R_{21,12}^{L,\text{eff}}(\omega_{21})\right)\\
 & =\sin^{2}(\delta)f^{2}(\epsilon)\Gamma_{\text{eff}}^{L}(\omega_{21})=\pi\sin^{2}(\delta)f^{2}(\epsilon)J_{\text{eff}}^{L}(2\tilde{\Delta})\big(n^{L}(2\tilde{\Delta})+1\big),
\end{aligned}
\end{equation}
and
\begin{equation}
\begin{aligned}
\text{Re}\left(R_{12,21}^{R,\text{eff}}(\omega_{12})\right) & =\text{Re}\left(R_{21,12}^{R,\text{eff}}(\omega_{12})\right)\\
 & =\cos^{2}(\delta)f^{2}(\epsilon)\Gamma_{\text{eff}}^{R}(\omega_{12})=\pi\cos^{2}(\delta)f^{2}(\epsilon)J_{\text{eff}}^{R}(2\tilde{\Delta})n^{R}(2\tilde{\Delta}),\\
\text{Re}\left(R_{12,21}^{R,\text{eff}}(\omega_{21})\right) & =\text{Re}\left(R_{21,12}^{R,\text{eff}}(\omega_{21})\right)\\
 & =\cos^{2}(\delta)f^{2}(\epsilon)\Gamma_{\text{eff}}^{R}(\omega_{21})=\pi\cos^{2}(\delta)f^{2}(\epsilon)J_{\text{eff}}^{R}(2\tilde{\Delta})\big(n^{R}(2\tilde{\Delta})+1\big).
\end{aligned}
\end{equation}
where we define $\epsilon\equiv\lambda/\Omega$. The resulting heat current is
\begin{equation}
\begin{aligned}j_{{\rm {NESS}}} & =-4\tilde{\Delta}\text{Re}\Big[R_{21,12}^{L,\text{eff}}(\omega_{21})\rho_{22}^{SS}-R_{12,21}^{L,\text{eff}}(\omega_{12})\rho_{11}^{SS}\Big]\\
%
%
 %
 & =\frac{4\pi\tilde{\Delta}\sin^{2}(\delta)\cos^{2}(\delta)f^{2}(\epsilon)J_{\text{eff}}^{L}(2\tilde{\Delta})f^{2}(\epsilon)J_{\text{eff}}^{R}(2\tilde{\Delta})\left(n^{L}(2\tilde{\Delta})-n^{R}(2\tilde{\Delta})\right)}{\sin^{2}(\delta)f^{2}(\epsilon)J_{\text{eff}}^{L}(2\tilde{\Delta})\big[2n^{L}(2\tilde{\Delta})+1\big]+\cos^{2}(\delta)f^{2}(\epsilon)J_{\text{eff}}^{R}(2\tilde{\Delta})\big[2n^{R}(2\tilde{\Delta})+1\big]}.
 \label{eq:Ajness3}
\end{aligned}
\end{equation}
\end{widetext}
We now interpret this result.
The Effective Hamiltonian heat current expression captures {\it only} sequential-resonant transport processes, similarly to the weak coupling limit. This is reflected by the trigonometric factors $\sin^2{\delta}$ and $\cos^2\delta$
that project the $\hat S^L$ and $\hat S^R$ coupling operators to the $\hat\sigma^x$ component at each side.
Thus, the current does not ``make use" of the coupling component into the $z$ direction. 
Another signature of sequential-resonant transport is that it takes place at the spin splitting frequency $2\tilde\Delta$, here dressed  by the coupling function to the bath according to Eq. (\ref{eq:h_rc_s_qubit_angle}).

The coupling energy $\lambda$ appears in several places: it modifies the gap $\Delta\to \tilde \Delta$. It also appears as a prefactor $\lambda^2$ within the spectral function. The dressing functions $f(\epsilon)$ with $\epsilon=\lambda/\Omega$ depend on $\lambda$ in a nontrivial manner. However the frequency that dictates transport is the spin splitting $\Delta$ rather than $\Omega$, as we observe in Fig.
\ref{fig:5}.
This highlights that the Effective Hamiltonian framework is not capable of capturing transport effects directly between the baths, which are dominant for noncommuting models at strong couplings~\cite{NicknonC}.

We also clearly see the effect of thermal rectification, even at identical coupling energies $\lambda_{L,R}$. It emerges due to the 
use of different operators at the two ends showing up here in the  distinct factors $\sin^2 \delta$ and $\cos^2 \delta$. In the denominator, these factors appear separately in front of the left and right bath thermal distributions, allowing current asymmetry upon temperature exchange. 

Finally, we comment that Eq. (\ref{eq:Ajness3}) properly reduces to the ultraweak coupling limit. In this case,  $\tilde \Delta \xrightarrow{\lambda/\Omega \to 0} \Delta$ and 
$J_{\text {eff}}^{\alpha}(\omega)\xrightarrow{\lambda/\Omega\to 0} J^{\alpha}(\omega)$ [see Eq. \eqref{eq:spec_fun_brownian}], recovering known results \cite{Segal:2005,Segal:2006}, albeit here with the coupling operators furthermore projected to their $\hat\sigma^x$ component through the 
$\sin \delta$ and $\cos \delta$ coefficients. 

\bibliography{bibliography}

\end{document}